# WRINKLED FLAMES AND GEOMETRICAL STRETCH

Bruno Denet[1*] and Guy Joulin[2#]

[1] Institut de Recherche sur les Phénomènes Hors d'Equilibre, UMR 6594 du CNRS,
Technopole de Château Gombert, 49 rue Joliot-Curie, 13384 Marseille Cedex 13, France.

[2] Institut P-prime, UPR 3346 du CNRS, ENSMA, Université de Poitiers,
1 rue Clément Ader, B.P. 40109, 86961 Futuroscope Cedex, Poitiers, France.

Localized wrinkles of thin premixed flames subject to hydrodynamic instability and geometrical stretch of uniform intensity ($S$) are studied. A stretch-affected nonlinear and nonlocal equation, derived from an inhomogeneous Michelson-Sivashinsky equation, is used as starting point and pole-decompositions as tool. Analytical and numerical descriptions of isolated (centred or multi-crested) wrinkles with steady shapes (in a frame) and various amplitudes are provided; their number increases rapidly with $1/S > 0$. A large constant $S > 0$ weakens or suppresses all localized wrinkles (the larger ones, the easier) whereas $S < 0$ strengthens them; oscillations of $S$ further restrict their existence domain. Self-similar evolutions of unstable many-crest patterns are obtained. A link between stretch, nonlinearity and instability with the cut-off size of wrinkles in turbulent flames is suggested. Open problems are evoked.

PACS number(s): 47.20.-k, 47.32.-y, 82.33.Vx

---

[*] e-mail : bruno.denet@irphe.univ-mrs.fr
[#] e-mail : guy.joulin@lcd.ensma.fr

# I. INTRODUCTION

Being of chemical and diffusive origins flame fronts propagating in premixed gases are markedly subsonic. Thus, because the fresh ($\rho_u$) and the burnt ($\rho_b < \rho_u$) gas densities differ, flame deformations almost instantly modify the piecewise-incompressible surrounding flow, and conversely. This *nonlocal* hydrodynamic flame/flow feedback brings about the Darrieus [1]-Landau [2] (DL) wrinkling instability, and greatly complicates the free-boundary (hence nonlinear) dynamical flame-front problems, especially from a theoretical view point.

In the limit of small Attwood numbers $0 < \mathcal{A} = (\rho_u - \rho_b)/(\rho_u + \rho_b) \ll 1$ the DL instability is weak, though. Sivashinsky's seminal work [3] showed how flat-on-average flames propagating into a quiescent pre-mixture then evolve according to a nonlocal nonlinear partial differential equation (PDE); subsequent works [4] essentially improved $\mathcal{A}$-dependent coefficients therein. Michelson's numerical early study [5] of the PDE showed that unforced flame fronts soon acquire the shape of parabola-like arches joined by sharper crests pointing towards the burnt gas; at late times a single steady arch (or half-one) with a maximum wavelength compatible with periodic (or Neumann) boundary conditions generically survives in (not-too-wide) channels.

Further numerical investigations [6] of this (Michelson-) Sivashinsky (MS) PDE, exploiting the pole-decomposition method [7], evidenced a much richer manifold of steady solutions whose number increases with available lateral size noticeably faster than linearly. Most of those happen to be unstable and cannot be evidenced by more conventional time-marching procedures.

It would be nice to understand whether a simple mechanism underlies such a proliferation of steady solutions, which has so far not been done. The present analyses suggest that a *geometrical stretch* ($S$) induced at large enough scale by a front curvature of the proper sign, can noticeably contribute: it indeed generates new couplings with the mechanisms already at work in the MS equation, and generates novel solutions that are steady in a suitable frame. Though unstable, these may play a nontrivial role, like the long-lived weakly unstable states of flat-on-average flames forced by weak noise [6]; or as building blocks of the travelling bursts that are randomly 'emitted' near troughs of wide flames [8]. Hence a side question: is it a spontaneous emission?

A related problem of great practical importance is to understand how velocity modulations or fluctuations in the incoming flow of fresh gas, *e.g.* time-dependent or even turbulent ones, affect the front dynamics. The problem is unfortunately much too tough for a frontal theoretical attack.

Flames could conceivably be passively deformed at high intensities of forcing, but even the statistics of this is not yet fully understood theoretically [9]. Moreover, the inner cut-off length of front wrinkles in turbulent flows is experimentally known [10] to coincide with the neutral wavelength at which DL instability and curvature affects balance, strongly indicating that the smallest detectable flame wrinkles are not passive.

This hint of forcing-instability interaction at the cut-off length motivates another question: how do incoming velocity *modulations* affect instability-driven patterns at or about such scales?

The point is conceptually different from that of *implantations* of incipient wrinkles by the *small*-scale components of a forcing; these are indeed often present, purposely or because most numerical integration methods are noisy. Despite early attempts about passive forced propagations [11][12] the related mechanism of complex singularity implantation is not understood; unfortunately, it ultimately has a strong impact on flame dynamics [13]. Pole-sprinkling [14] mimics implantations and somehow bypasses the problem but, although tried for expanding flames [15], this trick is still hampered by its computational cost. To study space- or angle-periodic forced fronts, pseudo-spectral integration remains more flexible and faster [16].

The present work has a more modest scope: it focuses on *already implanted* patterns and inquires about their long term viability in the presence of a *geometrical stretch* caused by an underlying curved front, itself possibly influenced by *large* scale incoming velocity modulations.

The paper is organised as follows. An extended MS equation accounting for geometrical stretch is introduced in Section II. Pole-decomposed steady solutions (in *a* frame) are presented in Sec. III, and specialized to a simple centred crest then to ones with larger amplitudes. The predictions are compared to numerical results in Sec. IV. Multi-crested 'steady' wrinkles are identified and analyzed in Sec.V, and their self-similar unsteady counterparts discussed in Sec.VI. Section VII takes up varying $S$s . Section VIII tentatively relates the cut-off scale of wrinkling to stretch, and Sec. IX gathers final remarks and open problems. Appendices A, B detail technical points.

**II. NONLINEAR EQUATION FOR STRETCHED ISOLATED WRINKLES**

The starting point adopted here to take up the aforementioned topics is a non-dimensional forced version of the (Michelson-) Sivashinsky [3] PDE:

$$\varphi_t + \tfrac{1}{2}\varphi_x^{\,2} = \nu \varphi_{xx} + I(\varphi, x) + u(t, x) \,. \qquad (2.1)$$

It governs the shape $y = \varphi(t, x)$ of a flame observed in the frame $(x, y)$ of a flat one ($\varphi \equiv 0$) propagating towards the fresh gases ($y < 0$) at the laminar flame speed $u_L$. The subscripts $t$ (time) and $x$ (abscissa) stand for partial derivatives, and the linear *nonlocal* operator $I(\varphi, x)$, defined by $I(e^{ikx}, x) = |k| e^{ikx}$, accounts for the hydrodynamic DL instability that affects premixed flames at long wavelengths. The scaled Markstein length [17] $\nu > 0$, proportional to the actual flame-front thickness $\ell$ based on $u_L$ and the fresh-gas heat diffusivity, measures how the local burning speed $u_n$ of a flame element relative to fresh gases responds to curvature, $u_n - u_L \sim \nu \varphi_{xx}$.

With $u(t, x) = 0$, the linearized Eq.(2.1) admits normal modes $\varphi \sim \exp(ikx + \varpi t)$ whose growth or decay rate is $\varpi(k) = |k|(1 - \nu |k|)$. In the present units [$\sim \mathcal{A}^{-2} \ell / u_L$ for time, $\sim \ell / \mathcal{A}$ for abscissa and Markstein length, $\sim \ell$ for wrinkle amplitude, $\sim \mathcal{A}^2 u_L$ for speed variations], the range of unstable wave-numbers and the shortest growth time are $0 \leq |k| \leq k_n = 1/\nu$ and $1/\varpi(k_n/2) = 4\nu$, respectively. The 'eikonal' term $\varphi_x^2/2$ in (2.1) results from the secant of the small angle $\alpha(t, x)$ between local (normal to the front) and mean ($y$-axis) directions of propagation, $(1 + s^2)^{1/2} - 1 = s^2/2 + \ldots$, $s = \tan(\alpha) \sim \mathcal{A} \varphi_x$. The stabilising effect of $\varphi_x^2/2$ can saturate the DL instability [5], soon leading to parabola-like arches when $u(t, x) = 0$.

The forcing term $u$ in Eq. (2.1) represents the $y$-wise component of a shear-flow type of modulation in the fresh-gas flow; $u(t, x) \neq 0$ may also result from an inhomogeneous and/or fluctuating fresh gas composition that makes $u_n$ change [18]. Without its DL contribution $I(\varphi, x)$, Eq.(2.1) would be an inhomogeneous Burgers equation modelling *passive* propagations.

Let $\Phi(t, x)$ represent a solution of Eq. (2.1). Another one, $\varphi(t, x)$, is sought in the form $\varphi = \Phi + \phi$, where $\phi(t, x)$ represents extra wrinkles grown on top of the base flame shape $\Phi$. The 'excess' function $\phi(t, x)$ satisfies the homogeneous equation

$$\phi_t + \Phi_x \phi_x + \tfrac{1}{2} \phi_x^2 = \nu \phi_{xx} + I(\phi, x), \tag{2.2}$$

where the multiplicative $\Phi_x$ accounts for the *geometry* of the base solution of Eq.(2.1), and hence is partly tributary of the forcing function in (2.1), especially if it is large. In fact, the structure of (2.1) indicates how to select $u(t, x)$ to yield (almost-) any presumed $\Phi(t, x)$.

We next assume that, possibly helped by a proper choice of the forcing function, one may approximate the base pattern $\Phi(x,t)$ by a parabola over the region where a localized $\phi_x$ 'lives'. With $\Phi_x \approx Sx + U$, Eq.(2.2) acquires the simpler, yet still nonlocal and nonlinear, form

$$\phi_t + (Sx + U)\phi_x + \tfrac{1}{2}\phi_x{}^2 = \nu \phi_{xx} + I(\phi, x), \qquad (2.3)$$

first proposed as a model with $U = 0$ and briefly studied in [19]. Whatever $U$ is, Eq.(2.3) admits infinitesimal solutions $\phi = A(t)\exp(iq(t)(x - \chi(t)))$, provided the wave-number $q(t)$, amplitude $A(t) << \lambda(t) = 2\pi / q$ and shift $\chi(t)$ satisfy $dq/dt = -Sq$, $dA/dt = A\varpi(q)$, $d\chi/dt = U$. So, while the disturbance drifts a positive $S > 0$ *stretches* its wavelength, $\lambda(t) \sim \exp(\int_0^t S(t')dt')$; $S < 0$ is often termed 'compression'. $U(t)$ can be formally removed by a change of coordinate ($x - \chi(t)$ used instead of $x$) and is henceforth omitted. Yet it must be kept in mind that *the 'steady' patterns encountered later will only be so in a specific frame drifting at the uniform lateral speed $U$*. A locally uniform $x$-wise gas velocity $v(t,0)$ and the associated $v(t,0)\phi_x$ could have been accounted for in Eq.(2.3) then lumped in $U$ and 'eliminated', but the $S$-dependent term cannot. The curvature $S$ of the base pattern $\Phi$ is here termed "stretch intensity". Incidentally, a $Sx\phi_x$ term appears when studying self-similar rational solutions to the Burgers equation [20]; $x = X/E$ is then an abscissa measured in a $t$-dependent unit of length $E(t)$.

### III. CENTRED STEADY PATTERNS

Attention is from now on focused on *finite-amplitude* solutions to Eq.(2.3), namely: *localized* patterns $\phi$ that are peaked near the origin, and have $\phi_x{}^2$ and $I(\phi, x)$ decaying like $(\nu/x)^2$ at large distances; the DL term $I(\phi, x)$ is then $\fint_{-\infty}^{+\infty} \phi_{x'}/(x - x')\pi \, dx'$. Time-independent $S$s, as exist about the troughs (wide local minima) of steady base flame profiles $\Phi$ when unforced [21], are considered first ($t$-dependent ones will be touched upon in Sec.VII). Localised solutions to the nonlinear (2.3), *e.g.* the steady ones (in a suitable frame) considered below, bring about a continuum of Fourier modes, which makes them difficult to handle numerically by spectral methods. The pole-decomposition technique recalled below bypasses the difficulty

## A. Pole equations

As first shown in [7][22], the MS PDE (Eq.(2.3) with $U = 0 = S$) admits exact non-periodic solutions $\phi$ representing localized patterns that are 'steady' and have

$$\phi_x = \sum_{n=-N}^{N} \frac{-2\nu}{x - iB_n}. \tag{3.1}$$

The $iB_n$s, with $B_n > 0$ for $n = 1, 2 \ldots N$ and $B_{-n} = -B_n$ ($\phi$ is real when $x$ is), are poles of the flame slope $\phi_x$ continued in the complex $z$-plane, $z = x + iB$. These carry the same 'charge' (residue) $-2\nu$, fixed by the dominant balance $\phi_x^2 / 2 \sim \nu \phi_{xx}$ near each $z = iB_n$. Equation (3.1) was shown in [19] to also hold when $S \neq 0$. When $t$-independent such $B_n$s, $|n| = 1, \ldots, N$, obey:

$$\sum_{n \neq m = -N}^{N} \frac{2\nu}{B_n - B_m} - \text{sgn}(B_n) + SB_n = 0, \tag{3.2}$$

where the sgn(.) function results from the DL instability [$1/(x - iB)$ has $-i\,\text{sgn}(B)/(x - iB)$ as Hilbert transform]; the sum comes from nonlinearity and $SB_n$ from stretch. Since $B_{-n} = -B_n$, summing Eqs.(3.2) over $n \geq 1$ yields $B_{bar} - SB_{rms}^2 \equiv \nu(2N - 1)$; this 'sum rule' exactly relates the barycentre (or centre of mass) $B_{bar} \equiv (B_1 + \ldots + B_N)/N$ of the positive $B$s to their variance $B_{rms}^2 \equiv (B_1^2 + \ldots + B_N^2)/N$, a useful check of accuracy for numerical resolutions.

## B. Elementary centred steady crest

The case of one pole pair ($N = 1$, $\phi_x - iI(\phi, x) = -4\nu/(x - iB_1)$, $B_1 > 0$), corresponding to an elementary steady wrinkle centred at $x = 0$, is particularly simple as (3.2) gives a quadratic:

$$\nu / B_1 - 1 + SB_1 = 0. \tag{3.3}$$

A single real solution $B_1 \leq \nu$ is found for $S \leq 0$ (compression), but two of them exist if $0 \leq S < S_c \equiv 1/4\nu$, $\nu \leq B^- \leq B^+$, with $B^-$ (or $B^+$) going to $\nu$ (or $+\infty$) as $S\nu \to 0^+$. Two 'steady' elementary crests are then admissible, that with $B_1 = B^-$ being sharper and narrower. If $0 < S\nu \ll 1$, the $B^-$ root essentially results from a balance between DL instability and nonlinearity, $1 \approx \nu / B^-$; $B^+$ is then of a different type mainly governed by DL effect and stretch, $1 \approx SB^+$. Both branches $B^\pm(\nu, S)$ merge at $B_c = 2B^-(\nu, 0) = 2\nu$ when $S$ coincides with the dimensionless maximum growth rate, $S_c = \varpi(k_n / 2)$ (see Sec. II); $2S_c B_c = 1$ then holds. No steady solution with a single pair of poles $\pm iB_1$ is allowed if $S > S_c$. The generalization of Eq.(3.3) to $t$-dependent $B$s (Sec. IV) reveals that the $B^+$ root is unstable, and that $S > S_c$ leads to $B_1(t/\nu \to \infty)/\nu = \infty$, i.e. wrinkle suppression.

Importantly, *the DL instability is a sine qua non of this stretch-induced crest suppression*, needed as it is to balance two stabilizing effects (see (3.3)): one is only intense at short scale (nonlinearity) and the other at large distance (stretch), and the balance is impossible if $S > S_c$.

### C. Large centred steady crests

In the stretch-free case, the uppermost pole altitude increases with the number $N$ of pole pairs, and the typical difference $B_n - B_{n-1}$ gets small compared to $B_{max} = \max(B_n)$ if $N \gg 1$; one may then replace the discrete sum featured in Eq.(3.2) by an integral over a continuous measure [7], such that $P(B)dB$ is the number of imaginary poles $iB$ with 'altitudes' between $B$ and $B + dB$. When applied to Eq.(3.2) the continuous approximation leads to an integral equation for the density $P(B) = P(-B)$:

$$\fint_{-B_{max}}^{+B_{max}} \frac{2\nu P(B')dB'}{B - B'} = \text{sgn}(B) - SB, \quad (3.4)$$

where the principal-part integral complies with the constraint $m \neq n$ in Eq.(3.2). Although (3.4) formally is the difference between its $S = 0$ version and a Wigner equation (no $\text{sgn}(B)$ in the right-hand side, [23]) one may not subtract partial solutions: (3.4) does not hold for $|B| > B_{max}$, where no pole lies, and $B_{max}$ itself must be found as part of the complete solution, thanks to the overall normalisation $\int_0^{B_{max}} P(B')dB' = N$.

Equation (3.4) is solved by a Fourier method like in [21] (see Appendix A). In terms of an angle $-\pi/2 \leq \theta \leq +\pi/2$ defined by $B = B_{max} \sin\theta$, $P(B)$ reads:

$$2\pi^2 \nu P(|B| \leq B_{max}) = \ln(\cot^2(\theta/2)) - \pi S B_{max} \cos\theta, \quad (3.5)$$

and $P(|B| \geq B_{max}) = 0$. With $\sin\theta = B/B_{max}$ and $2\nu P(B)$ fixed by Eq.(3.5), a contour integration in $\theta$-plane (or p.591 of [24]) expresses $\phi_x = -\int_{-B_{max}}^{+B_{max}} 2\nu P(B')dB'/(x - iB')$ in terms of $\sinh\xi \equiv x/B_{max}$ as $\text{sgn}(-x)\pi\phi_x = \ln(\coth^2(\xi/2)) - SB_{max}\cosh\xi$. The ensuing crest profile reads $\text{sgn}(x)\pi\phi(x)/B_{max} = -\sinh\xi \ln[\coth^2(\xi/2)] - 2\xi + SB_{max}(\xi + \sinh\xi \cosh\xi)/2$, up to an additive constant. The large steady crest thus has $(\phi(x)-\phi(0))/B_{max} = F(|x|/B_{max}, SB_{max})$, and is $\nu$-independent. The cumulative pole distribution $R(B) \equiv \int_0^B P(B')dB'$ deduced from (3.5) is:

$$2\pi^2 \nu R(B)/B_{max} = \sin\theta \ln[\cot^2(\theta/2)] + 2\theta - SB_{max}(\theta + \sin\theta\cos\theta)/2, \qquad (3.6)$$

to be compared with $i\pi\phi(ix)$. The corresponding centre of mass of positive $B$s, $B_{bar} \approx \int_0^{B_{max}} P(B')B'dB'/N$, has $\pi B_{bar}/B_{max} = (1-w/3)/(1-w/4)$, $w \equiv \pi S B_{max}$, and can be notably less than the yet unknown $B_{max}$; the 'shape factor' $B_{rms}/B_{bar}$ also varies weakly with $w$. Finally, the normalisation of $P(B)$ re-written as $R(\theta = \pi/2) = N$ completes the resolution of Eqs.(3.4). It relates $B_{max}$ and $\nu N$ to the stretch intensity $S$ by an analogue of (3.3):

$$(2\pi N\nu)/B_{max} - 1 + (\pi S/4)B_{max} = 0, \qquad (3.7)$$

the structure of which again parallels (3.2) and the 'sum rule' (expressed as $(2N-1)\nu/B_{bar} - 1 + (B_{rms}/B_{bar})^2 SB_{bar} = 0$).

A single large crest again exists at fixed $N\nu$ whatever $S \leq 0$ is, and has $B_{max} \leq 2\pi\nu N$. Just like with (3.3) two values $B^{\pm}$ of $B_{max}$ (hence two densities $P^-(B)$, $P^+(B)$) are obtained for $0 \leq S \leq S^*$, namely $2\pi\nu N \leq B^-(\nu N, S) \leq B^*$ and $B^+(\nu N, S) \geq B^*$ with:

$$S^* \equiv 1/(2\pi^2 N\nu), \quad B^* \equiv 4\pi N\nu. \qquad (3.8)$$

Similar to the 2-pole case, both branches merge at $B^* = 2B^-(\nu N, 0)$ when $S = S^*$. No real $B_{max}$ exists if $S > S^*$, and the DL mechanism (middle term of (3.7)) again is needed for crest suppression. And *larger wrinkles are easier to suppress*: $\nu S^* \sim 1/N$.

Put differently, a given stretch intensity $0 < S\nu \ll 1$ allows for a large centred steady crest iff $1 \ll N \leq N^* \equiv \lfloor 1/2\pi^2 \nu S \rfloor$ ($\lfloor . \rfloor \equiv$ integer-part). In case the number of pole pairs exceeds $N^*$ initially, at least $N - N^*$ of them will ultimately be expelled to $|B| \gg N\nu$ as time elapses (see below for the pole dynamics): *the stretch intensity selects the width and amplitude of the surviving crest*; *both scale like the radius of curvature $1/S$ of the base flame* in any case : $2S^*B^* = 4/\pi \approx 1.27$ for $N \gg 1$, $2S_c B_c = 1.00$ for $N = 1$.

## IV. NUMERICAL *vs.* ANALYTICAL

Two numerical approaches were employed to study Eqs.(3.2): (i) use Newton iterations or kin, that are delicate to initiate in case of multiple solutions yet give access to stable and unstable ones, and can benefit of analytically determined seeds; (ii) acknowledge that equations (3.2) are the restriction to steady and pure imaginary poles $iB_n$ of more general equations [7][19] for the poles $z_j(t) = x_j(t) + iB_j(t)$ of $\phi_x$ in unsteady situations, *viz.*:

$$\frac{dz_j}{dt} = \sum_{-N, q \neq j}^{+N} \frac{2\nu}{z_q - z_j} - i\,\mathrm{sgn}(\Im(z_j)) + Sz_j, \qquad (4.1)$$

where $\Im(.)$ denotes an imaginary part; the front slope $\phi_x(t,x)$ then is a sum, similar to Eq.(3.1), of $-2\nu/(x - z_j(t))$ contributions. Each pole pair constitutes a soliton and, as the dynamics (4.1) conserves their number $N$ (if finite), the pole-decomposition method is noise-free. Numerically integrating Eqs.(4.1) only yields stable equilibriums at large times, but gives access to stability properties of any steady solution already at hand.

When restricted to $z_j = iB_j$, the numerical procedure(s) always led to an equilibrium if $N < N^*$ and the pole imaginary parts initially have $|B| < B^+$. The numerical pole density $P_{num}$, defined by $P_{num}((B_j + B_{j-1})/2) \equiv 1/(B_j - B_{j-1})$ and linear interpolation in between, is compared with the predictions (3.5)(3.7) in Fig.1 for $1/\nu = 199.5$, $N = 100$, $S = 0.05 \approx S^*/2$, showing excellent agreement ... up to $B = (B_N + B_{N-1})/2$; this also holds for the corresponding cumulative density $R(B)$ along the lower branch [*i.e.*, with $B_{max} = B^-(\nu N, S)$ in Eq.(3.5)], again not too close to $|B| = B_{max}$, a point to be commented at the end of this section.

Still, in no way could one obtain a density $P_{num}(B)$ resembling Eq.(3.5) if used with $B_{max} = B^+(\nu N, S)$, even when $B^+(\nu N, S)$ and $B^-(\nu N, S)$ have comparable magnitudes. This relates to an additional constraint, not visible in (3.4) (3.7): pole densities must be non-negative. For $|B| \approx B_{max}$ the analytical prediction (3.5) has $2\pi^2 \nu P(B) \approx (2 - \pi S B_{max})\cos\theta$, which gets negative indeed at $|B| \lesssim B_{max}$ if $SB_{max} > 2/\pi = S^* B^*$, see Eq.(3.7). Thus, even though our starting assumptions $N \gg 1$ and $S < S^*$ are fulfilled, the upper branch $B_{max} = B^+(\nu N, S) > S^* B^*/S$ and the corresponding profile $\phi(x)$ are *spurious*.

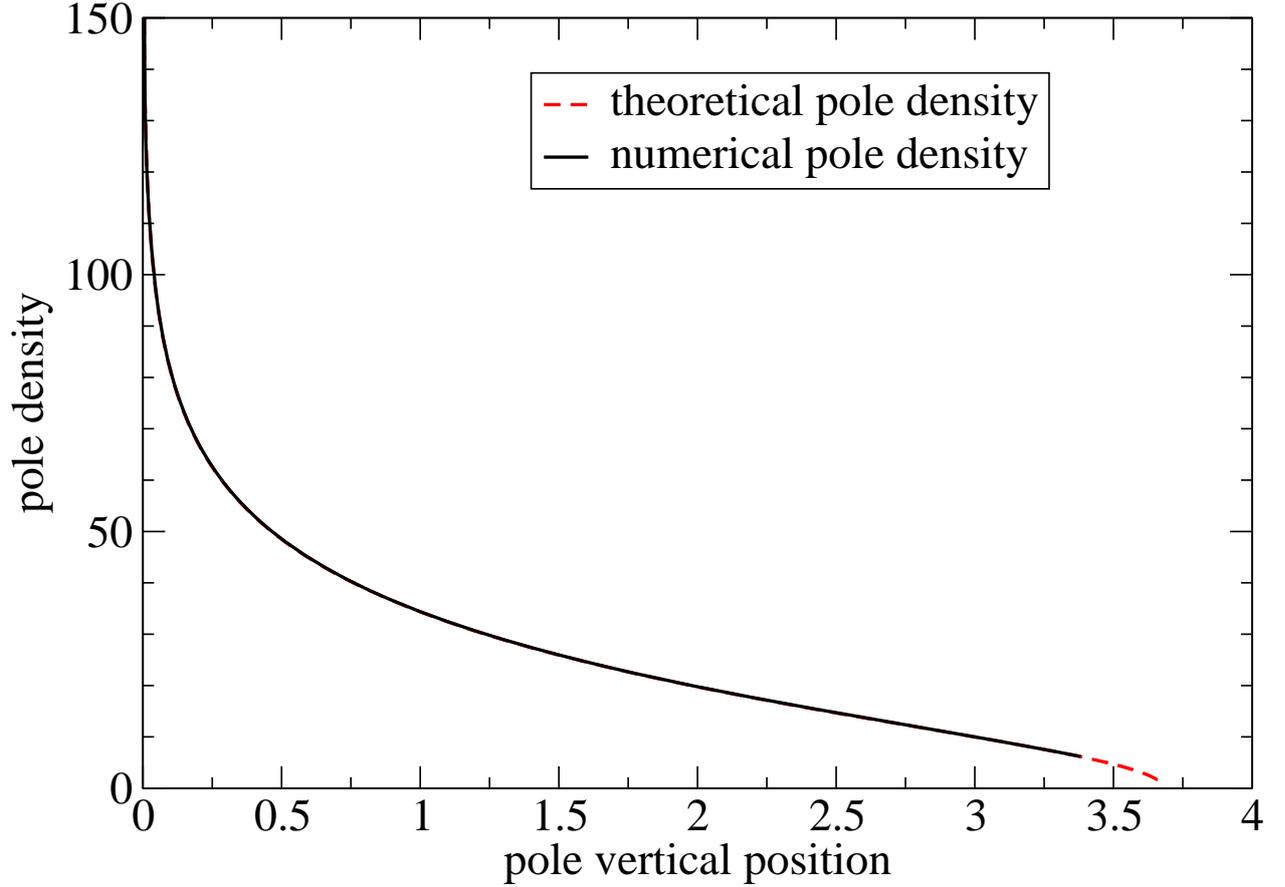

FIG. 1 (Color online) Numerical pole density $P_{num}(B)$ vs. pole altitude $B$ (solid black), compared to the analytical prediction $P^-(B)$, Eqs.(3.5) and (3.7) with $B_{max} = B^-(\nu N, S)$ (dashed red). Both curves have $1/\nu = 199.5$, $N = 100$ ($S* = 0.102...$) and $S = 0.05$.

Instead of $P^+(B)$ the density $P_{num}(B)$ obtained by Newton-Raphson iterations with $N$ pole pairs and $S < S*$ approached the prediction (3.5) if used with $B_{max} = B^-(\nu N, S)$; *or*, depending on the initial $B$s used as iteration seeds, the converged density $P_{num}(B)$ happened to be close to that corresponding to $B^-(\nu(N-1), S)$ but supplemented with a single pair of remote poles $\pm ih$, with $h \approx 1/S - \nu(4N-3) + ...$ for $0 < SN\nu \ll 1$. The latter estimate results from a balance (at $B = h$) between stretch, the DL effect, and the combined vertical repulsions from the complex conjugate (located at $B = -h$) and from the $2(N-1)$ other poles considered positioned at their barycentre ($B = 0$):

$$1 - Sh \approx \nu/h + 4\nu(N-1)/h, \quad 0 < SN\nu \ll 1, \qquad (4.2)$$

to be compared with (3.3). A more complete determination of $h$ accounts for the full pole distribution with density $P^-(B)$ spread over $[-B_{max},+B_{max}]$, instead of a mere global charge $4\nu(N-1)$ sat at $B=0$. Once the integrals over $P^-(B)dB$ are analytically evaluated (see pp. 591 and 393 of [24]) the equation for $h$ looks like (4.2), except for its last term that is replaced by $(2/\pi)\arcsin(B_{max}/h) - SB_{max}^2/(h+(h^2-B_{max}^2)^{1/2})$, with $B_{max} = B^-(\nu(N-1),S)$. This resumes the form (4.2) when $SN\nu \ll 1$ and provides one with a useful test of the numerical method and convenient seeds for iterations: for $S = 0.01$, $\nu = 0.1$ and $N = 4$, the more complete expression gives $h = 98.682\ 569$ while the exact (numerical) value is 98.682 595. The 'centre of charge' estimate (4.2) gives 98.682 645 and is still a few percent accurate up to nearly $S = S*$; this accuracy is to be used in Sec.VI .

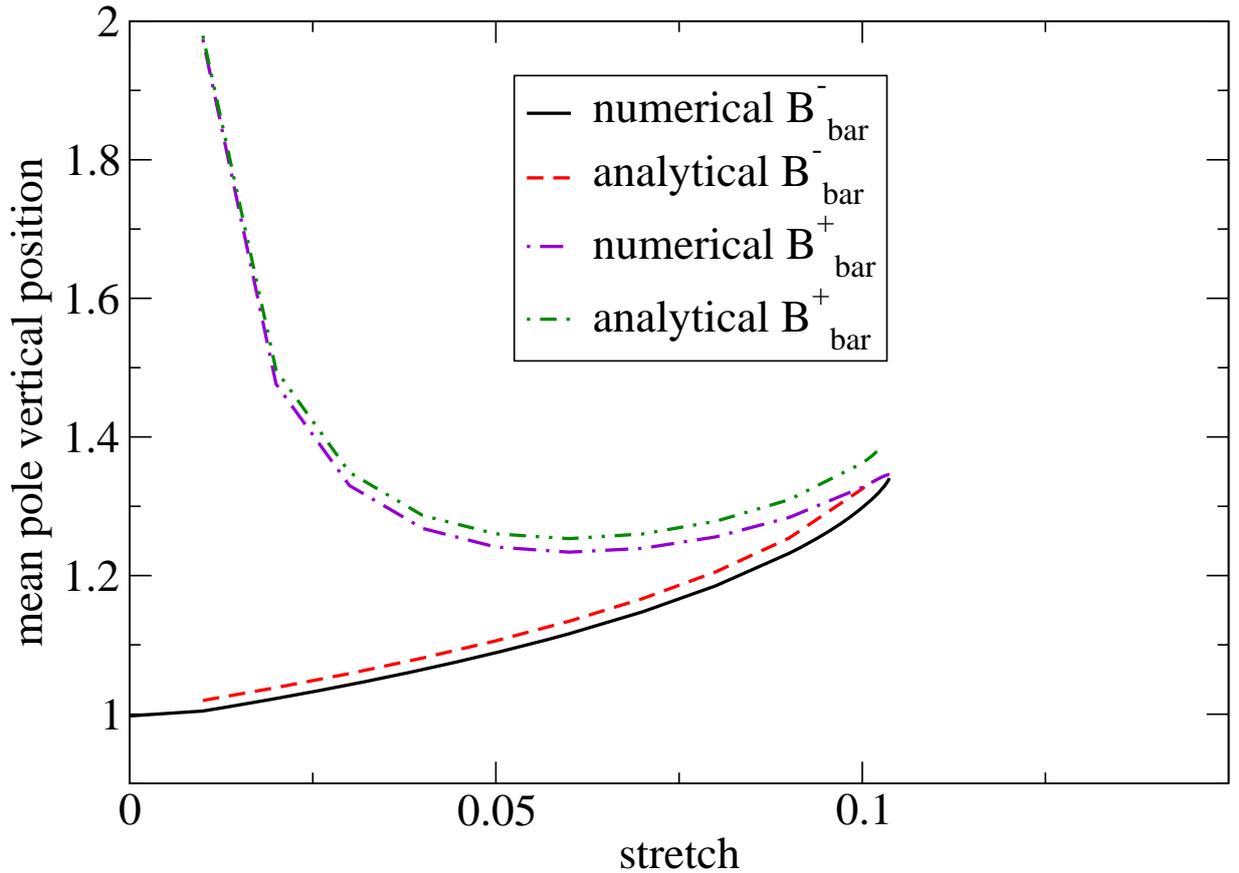

FIG.2 (Color online) Barycentres of $N = 100$ pole pairs aligned along the imaginary axis, *vs.* stretch intensity S, for $1/\nu = 199.5$ ($S* = 0.102...$, Eq.(3.8)), without detached poles (lower curves) or with them (upper curves). Solid (black) and dash-dot (violet) lines: Newton-Raphson iterations from the steady Eq.(4.1); Dash (red) and dash-doubledot (green) lines: analytical predictions see text.

Interestingly, the existence of a detached pole pair in equilibrium signals the appearance of a *new type of crest structures* that bifurcate from $|B|=\infty$ at $S=0^+$, in a sense generalizing the $B^+$ root of Sec.III (Eqs. (4.2) and (3.3) coincide if $N=1$). More general arrangements involving several remote poles will be encountered in sections V.B and VI.B.

Solving Eqs.(3.2) iteratively for different stretch intensities with a fairly large $N$ (=100) and $\nu \approx 1/2N$ gives the relationship between $S$ and the upper barycentre $B_{bar}$ shown in Fig.2; the modification of $B_{bar}(\nu N, S)$ caused by the existence of a detached pair of remote poles $\pm ih$ is also plotted and compared to the theoretical estimate $B_{bar}^+ = (1-1/N)B_{bar}(\nu(N-1),S)+h/N$. As expected from Eq.(3.7), the existence of a maximum admissible stretch intensity $S*$ [hence of a maximum admissible number $N*$ of aligned poles at fixed $S>0$] is numerically confirmed for large centred crests as well. However, while displaying similar trends to the corresponding numerical $B_{bar}$ and $B_{bar}^+$, both analytical predictions systematically exceed them by a visible amount that slowly increases with $S$ (and $N$). This in fact results from a failure of the continuous approximation wherever the density $P^-(B)$ gets too small, i.e. at $|B| \lesssim B_{max}$. As shown in [7] [21], good estimates of the *discrete* pole locations $B_j > 0$ can be retrieved from the continuous cumulative distribution $R(B)$ on solving $R(B_j) = j - 1/2$, $j = 1,...N$, but this deteriorates at $|B_j| \lesssim B_{max}$. Like for the extreme zeros of Hermite polynomials at large degrees [25], the region where $R(B)$ ceases to correctly count the pole labels has a size that may be estimated by $0 < R(B_{max}) - R(B_N) \sim 1$. With $N - R(B \lesssim B_{max})$ deduced from (3.6), this results in $B_N = B_{max}[1-(g(\pi S B_{max})/N)^{2/3}]$, $g(w) \sim (1-w/4)/(1-w/2) > 0$. Thus, $B_{max}/B_N - 1 > 0$ decays with $1/N$ and increases with $S$ (yet rather slowly), but $B_{max} - B_N \sim \nu(g(w)N)^{1/3}$ grows with $N$ and $S$: there is no numerical flaw in Figs. 1 and 2. Overestimating $B_N$ as $B_{max}$ in turn shifts $B_{bar}$ and $B_{bar}^+$ by about the same fractional amount.

## V. A VARIETY OF STEADY LOCALIZED PATTERNS

The previous centred isolated crests are unstable to lateral shifts. Writing the poles locations as $z_j(t) = D(t) + Z_j(t)$ leaves Eq.(4.1) invariant if $D(t) = D(0)\exp(St)$, whereby a steady crest pattern can be dragged as a whole by a geometry-induced tangential velocity $\Phi_x \approx Sx$, away from the base flame trough $x = 0$. Besides, [7] identified the reason why a population of nearby poles tends to align along parallels to the imaginary $z$-axis (see also [20]) and to build up a crest. This results from a 'horizontal' attraction encoded in the pair-wise interaction terms of Eqs.(4.1), combined with the 'vertical' repulsion featured in Eq.(3.2). If $|z_q - z_j| << B_{\max}$, $d(z_q - z_j)/dt \approx -4\nu/(z_q - z_j)$, whereby $\Im(z_q - z_j)\Re(z_q - z_j) \approx const.$ : $|\Re(z_q - z_j)|$ decays during the near collision, while $|\Im(z_q - z_j)|$ grows; this short-range alignment mechanism still operates with a non-zero $S$ [incidentally, it would still act at short distance in the presence of an extra damping in (2.2), say $\sim \phi$, but not at large scale].

### A. Nearly real crests/poles

One may thus conceive that steady twin crests might stay in equilibrium 'at' $x = \pm X_1$ ($X_1 > 0$) under two antagonistic actions: 'repulsive' lateral convection $SX$ ($S > 0$) and the horizontal attraction felt by each crest's pole, $+X_1 + ib_n$ say, caused by the $2N_1$ poles $-X_1 + ib_m$ belonging to the crest's twin. Balancing the two effects gives $X_1 \sim (2N_1\nu/S)^{1/2}$: if $S\nu N_1 << 1$, the pole-pile height $B_{\max} \sim 2N_1\nu$ is small compared to crest spacing $2X_1$ (as presumed), a situation henceforth referred to as 'nearly real' crests (or pole) arrangements.

For $N_1 >> 1$, $\nu N_1 = \mathcal{O}(1)$ and any $S$, the poles of a steady symmetric 2-crest pattern are not exactly aligned vertically but actually reside along two disjoint *curves*, $z(b) = x(b) + ib$ and $-x(b) + ib$, with $0 < x(b) = x(-b)$ and $|b| < b_{\max}$. Both curves share the pole density $p(b) = p(-b)$ per unit length along the $b$-axis, with $\int_0^{b_{\max}} p(b)db = N_1$ for normalization, and the real unknowns $x(b)$ and $p(b)$ obey a *complex*-valued generalization of (3.4) deduced from (3.2):

$$\fint_{-b_{\max}}^{+b_{\max}} \frac{2\nu p(b')db'}{z(b) - z(b')} = Sz(b) - i\,\text{sgn}(b) - \int_{-b_{\max}}^{+b_{\max}} \frac{2\nu p(b')db'}{z(b) + z(b')}. \tag{5.1}$$

Only for small stretch intensities, $0 < S\nu N_1 << 1$, could we solve (5.1) analytically for $z(b)$: to two orders all the poles are found to remain aligned at $\pm X_1 \pm ib + o(1)$, with $X_1 >> b = \mathcal{O}(\nu N_1)$ now defined by $SX_1 \equiv (2N_1.2\nu)/(2X_1)$. Their density $p(b)$ still obeys (3.4) with $\{p(b), N_1\}$ in lieu of $\{P(B), N\}$, but with $S$ replaced by $S + S/2$: as the $z(b)$ are not exactly real, the large-scale influence they feel from the other distant crest is not quite uniform, yet it may be Taylor-expanded for $|b| \ll 2X_1$ to produce a linear term $(z-X_1)(2N_1.2\nu)/(2X_1)^2 = (z-X_1)S/2$ that ultimately contributes the extra $S/2$. As net consequence the results for isolated crests, *e.g.* those encoded in Eqs.(3.5) (3.7) for $N \gg 1$, still hold for twin crests once $0 < S \ll 1/\nu N_1$ is replaced by $3S/2$; this replacement is required whatever $N_1$ is. If $0 < S\nu N_1$ gets $\mathcal{O}(1)$ the vertical pole alignments deform, with $|x(b)|$ decreasing faster at smaller $|b|$.

Multi-crested steady solutions also are unstable against shifts ($\sim \exp(St)$) as a whole. They then provide one with localized burst-like disturbances, akin to those invoked by Zel'dovich *et al.* [26] but here of *finite* amplitudes, *travelling along the nearly-parabolic base flame front* $Sx^2/2$; Fig. 3 shows a sample travelling burst comprising 8-poles. These bursts admittedly are also unstable with respect to modifications of the crest mutual distances, yet the presence of several nearby crests may help some survive longer as a result of the 'horizontal' interactions: to wit, many-crested pattern may be steady if properly centred (in the frame evoked below (2.3)).

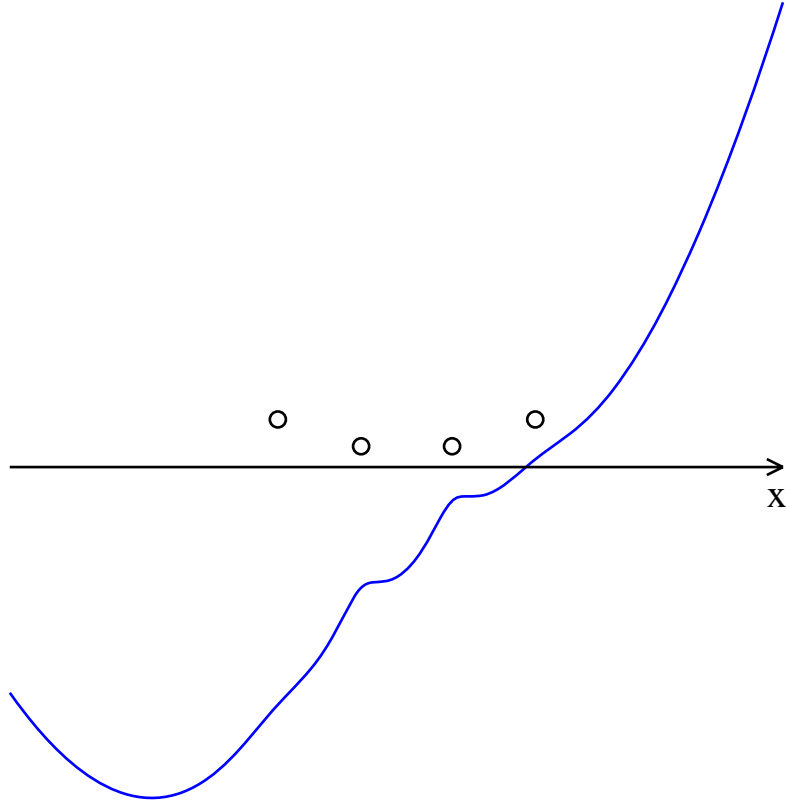

FIG.3 (Color online) Solid (blue) line: Snapshot of a finite-amplitude, 4-pole-pair lateral "burst" $\phi(t,x)$ superimposed to the base shape $Sx^2/2$ ($S=1, \nu=1/10, N=4$). The 4 pole locations in upper $x+iB$ plane are shown as open circles above the $x$-axis. The scale is identical in all directions

One might indeed have assumed that the previously considered isolated crest centred at $x = X_0 = 0$, with its $N_0$ pairs of poles $iB_{0,n}$, still is present: a steady 3-crest pattern is thus conceivable, with the lateral ones (again with $N_1$ pole pairs each) staying at a distance $\pm X_1$ given by $SX_1 = (2N_0.2\nu)/X_1 + (2N_1.2\nu)/(2X_1)$ if $0 < \nu SN_{0,1} \ll 1$. The construction may be pursued with 5 crests peaked at the abscissas $0 = X_0, \pm X_1, \pm X_2$, or 4 crests at $\pm X_1, \pm X_2$, ... etc. More generally, let $M$ be the number of crests involved in a steady arrangement and $N_k$ the number of pole pairs belonging to the $k$-th crest. This kind of configuration will henceforth be referred to as a ($N_1,...,N_k,...N_M$)-solution; and, whenever needed, the superscript "+" (e.g., $N_k^+$) will indicate that a pair of poles separates from the considered pole pile as $S\nu N_k \to 0+$. For $S > 0$ small enough that the typical crest spacing (anticipated to be $\sim (N_k \nu S)^{1/2}$) noticeably exceeds $(B_{\max})_k \sim \nu N_k$, the crest abscissas $X_k$ must satisfy the conditions of equilibrium of $M$ 'charges' $4\nu N_k$, $k = 1,...,M$, subject to attractive horizontal $1/X$ interactions and all sat on a common quadratic potential barrier $-SX^2/2$:

$$-\sum_{1=j\neq k}^{j=M} \frac{4\nu N_j}{X_k - X_j} + SX_k = 0. \quad (5.2)$$

Beside the condition $N_1 X_1 + ... + N_M X_M = 0$ that necessarily holds, the above crest number $M$ and weights $N_k$ are only constrained by $1 \leq M \leq N_{tot} \equiv N_1 + ... + N_M$. As a result, the combinatorial explosion makes the number $\mathcal{N}(N_{tot})$ of such conceivable "stretch *vs.* nonlinearity" nearly-real equilibriums rapidly grow with $N_{tot}$. The function $\mathcal{N}(N_{tot})$ grows faster than the number ($\sim \exp(\pi(2N_{tot}/3)^{1/2})/N_{tot}$ for $N_{tot} \gg 1$, [27]) of unordered integer partitions of $N_{tot}$ because unequal weights $N_k$ may be permutated; yet $\mathcal{N}$ grows less rapidly than that ($= 2^{(N_{tot}-1)}$, [28]) of ordered partitions because $x \leftrightarrow -x$ mirror images of admissible asymmetric patterns also are, which leads to double-counting: since most patterns are asymmetric, we conjecture that $\mathcal{N}(N_{tot}) \approx 2^{(N_{tot}-2)} \gg N_{tot}$ different patterns with near-real poles exist for $N_{tot} \gg 1$. At any rate, the above combinatorial reasoning suggests that *weak positive stretch allows for a proliferation of steady solutions.*

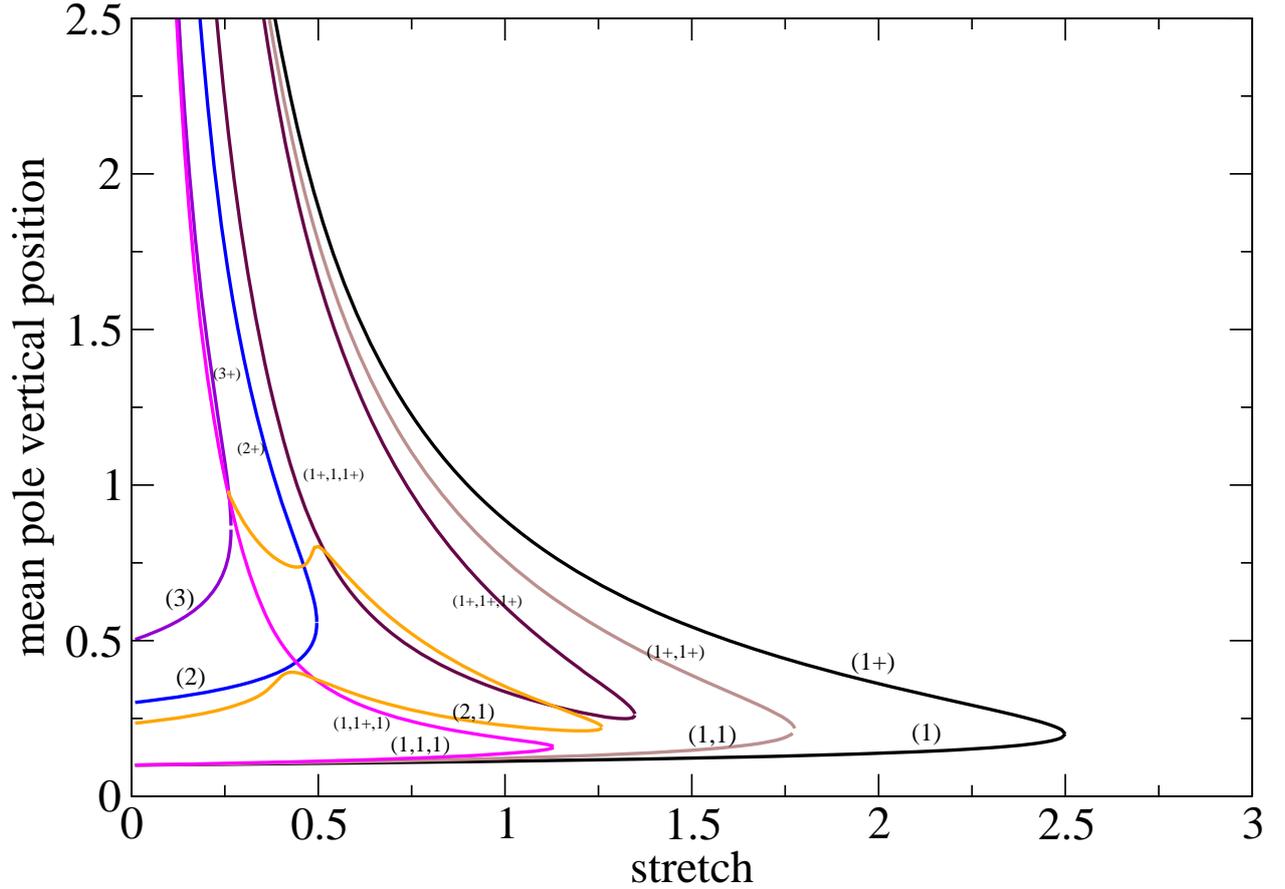

FIG.4 (Color online) A few curves giving $B_{bar}$ (barycentre of positive pole altitudes $B_k$ ) vs. $S$ (stretch intensity) at fixed $\nu = 1/10$, for pole arrangements with $N \leq 3$. The label arguments encode the number of near-vertical pole alignments (at small $S\nu$) and the number of pole pairs in each; the superscript "+" indicates the presence of a pair of remote poles, $|\Im(z)| \to \infty$ as $S\nu \to 0^+$: e.g., $(1^+, 1, 1^+)$ .

Only in the two particular instances evoked below could Eqs.(5.2) be solved analytically with unequal weights $N_k$.

The first situation has only two crests of arbitrary weights $N_-$ and $N_+$ located at $X_- < 0 < X_+$. Equations (5.2) readily yield $X_\pm = \pm 2 N_\mp (\nu / S(N_+ + N_-))^{1/2}$. A Taylor expansion like below (5.1), here for $|z - X_\pm| \ll X_+ - X_-$, shows that the poles belonging to either crest again receive an extra contribution from its neighbour, besides the imposed $0 < \nu S \ll 1$. Here, this entails the replacements $S \to S_\pm \equiv S(1 + N_\pm / (N_+ + N_-) + ...)$ for the stretch intensity effectively felt by either crest, all poles of which are aligned vertically within $o(\nu N_\pm)$ errors if $0 < \nu S N_\pm \ll 1$ and could thus be analysed via (3.2) or (3.4); $N_+ = N_-$ clearly yields $S_+ = 3S/2 = S_-$ like below (5.1). The more populated crest feels the higher effective influence of stretch and will the first to loose remote poles if $S\nu N_{\pm} > 0$ gets too high.

The second situation amenable to some analysis corresponds to a centred crest composed of $N_0$ vertically aligned pole pairs, symmetrically flanked by $N + N$ lateral crests that have $N_{|k| \geq 1} = N_1$ pole pairs each. The (horizontal) equilibrium of the central crest at $X_0 = 0$ is then guaranteed by symmetry. In the limit $0 < S\nu N_1 \ll 1$ the $N + N$ lateral crest locations $X_k = -X_k$, with $|k| = 1,...,N$, obey a particular form of (5.2):

$$\sum_{-N=j \neq k, 0}^{j=N} \frac{4\nu N_1}{X_k - X_j} + 4\nu \frac{N_0}{X_k} = S X_k \ . \qquad (5.3)$$

Its solutions can be expressed as $X_k = \text{sgn}(k)(4\nu N_1 \eta_{|k|} / S)^{1/2}$ in terms of the zeros $\eta_{|k|} > 0$ of the associated Laguerre polynomial $\mathcal{L}_N^{(\alpha)}(\eta)$ of degree $N$ and order $\alpha = N_0 / N_1 - 1/2$; this follows from Stieltjes' classical analysis of electrostatic problems [29] (see Appendix B).

For $N_0 = 0$ or $N_0 = N_1$ the $X_k$s actually are zeros of the $M$-th order Hermite polynomial, $\mathcal{H}_M(X_k(S/4N_1\nu)^{1/2}) = 0$, with $M = 2N$ or $2N+1$, since $\mathcal{H}_{2N}(\xi) \sim \mathcal{L}_N^{(-1/2)}(\xi^2)$ and $\mathcal{H}_{2N+1}(\xi) \sim \xi \mathcal{L}_N^{(+1/2)}(\xi^2)$ ([24], p.1001). Numerical resolutions of (4.1) in this case reveal that the "Hermite" approximation is excellent if $S\nu N_1 \ll 1$; but Table I shows it still works fairly well even for stretch intensities $S$ comparable to the value (again noted $S*$) at turning point.

| $|k|$ | Numerical $X_k$ s | Hermite $X_k$ s |
|---|---|---|
| 1 | 1.0281177 | 1.03253157 |
| 2 | 2.1081804 | 2.11689397 |
| 3 | 3.3416002 | 3.35449526 |

Table I: Numerically determined abscissas $X_k > 0$ of crests with "near-real" poles, *vs.* those estimated from the Hermite polynomial, $\mathcal{H}_M(X_k(S/4N_1\nu)^{1/2}) = 0$, for $N_k = N_1 = 1$, $M = 7$, $\nu = 1/10$ (this solution's turning point is at $S* \approx 0.55$), and $S = 0.25$. Like when $S\nu N_1 \ll 1$, all numerical poles have $|\Im(z_k)| \approx \nu$.

Table II shows similar results corresponding to $2N+1=7$, with $N_0=3$ pairs for the central crest and $N_1=1$ for the lateral ones: the fuller "Laguerre" approximation, still excellent at small $S\nu > 0$, again locates the crests fairly accurately for $S$s comparable to the value $S^*$ at turning point . Increasing the stretch intensity beyond $S^*$ makes a pole pair separate from the central pile, reducing its weight. For $0 < \nu S N_{0,1} \ll 1$ the effective stretch intensity felt by the central crest is $S_0 = S(1+1/\eta_1+...+1/\eta_N+...)$; as shown in Appendix B, $S_0$ also reads $S(1+2N/(N_0/N_1+1/2)+...)$ and may exceed the maximum stretch intensity that an isolated crest of weight $N_0$ can resist, $S^* \approx 1/(2\pi^2 N_0 \nu)$ for $N_0 \gg 1$, even if $\nu N_1 S \ll 1$.

| $|k|$ | Numerical $X_k$ s | Laguerre $X_k$ s |
|---|---|---|
| 1 | 3.4237352 | 3.48790272 |
| 2 | 5.6222914 | 5.6649485 |
| 3 | 8.0750858 | 8.10819913 |

Table II: Comparison between the numerically determined abscissas $X_k$ of $2N=6$ crests with "near-real" poles ($N_1=1$ pole pair each) in the presence of a central crest comprising $N_0=3$ pole pairs, and those ($X_k = \text{sgn}(k)(4\nu N_1 \eta_{|k|}/S)^{1/2}$) estimated from roots of the Laguerre polynomial $\mathcal{L}_N^{(\alpha=5/2)}(\eta)$ , for $\nu=1/10$ ( turning point at $S^* \approx 0.12$), and $S=0.06$.

Thus, the antagonistic actions of horizontal attraction between nearby crests and geometry-induced positive stretch near *any* flame trough (e.g., that of the base solution $\Phi$ we started from) generates new equilibrium positions: other poles may sit there if the effective stretch intensity they feel is compatible with a vertical equilibrium, making secondary troughs appear on the flame profile…and so forth. This gives hints on how the *stretch/nonlinearity competition is sufficient to contribute a complicated web of steady solutions to the MS equation* [6]. As shown below, still more exotic configurations exist, for nearly the same reason.

## B. Remote poles

In the above equilibriums between 'almost real' pole pairs in the presence of a weak enough stretch effect, Eq.(5.2), the DL instability mechanism little influence the crest- spacing directly, merely ensuring that nearly real poles remain so. Similar configurations could conceivably exist when the members of $N$ poles pairs are markedly off the real axis, provided these lie at nearly the same altitudes $\pm ih$ ; an elementary configuration of this type (a single detached pair) was encountered in Sec. IV. The height $h > 0$ has to significantly exceed $x_{max}$ for this to be viable: the repulsive influence that each of the $N$ poles $\approx x_k \pm ih$ inside one single row feels from the complex conjugates (now at a distance of $2h$) must indeed be nearly uniform, and weak enough not to destroy the possibility of a "DL vs. stretch" balance with $0 < SN\nu << 1$. For $h >> x_{max}$ the 'vertical' equilibrium requires

$$1 - Sh = 2\nu N / 2h + ... \tag{5.4}$$

at the two leading orders (compare to Eqs.(3.3)(4.2)), whereby $h = 1/S - \nu N + ...$ will self-consistently exceed $x_{max} \sim (\nu N / S)^{1/2}$ when $0 < S\nu N << 1$. As for the 'horizontal' equilibriums among the remote poles located at $z \approx x_k \pm ih$, one can show that their abscissas $x_k$s satisfy:

$$-\sum_{0=j\neq k}^{j=N} \frac{2\nu}{x_k - x_j} + Sx_k = 0, \quad k = 1,...,N, \tag{5.5}$$

to leading order for small positive $S$s. The Stieltjes analysis [29][35] implies the $x_k$s are again given by roots of a Hermite polynomial in first approximation, yet with a scale different from the nearly real cases: $\mathcal{H}_N(x_k(S/2\nu)^{1/2}) = 0$ ; since the largest zero of $\mathcal{H}_N$ is $\mathcal{O}(N^{1/2})$ whatever $N \geq 1$ is, $x_{max} \sim (\nu N / S)^{1/2} << 1/S$ if $N << 1/\nu S$ , as guessed. Though *a priori* limited to small stretch intensities (it is then excellent) this calculation evidences that *steady remote pole arrangements of a novel type exist when* $0 < S\nu << 1$ . Used with $B_k = \pm i(1/S - \nu N)$ and $\mathcal{H}_N(x_k(S/2\nu)^{1/2}) = 0$ as seeds for the pole locations, Newton-Raphson numerical resolutions of the steady Eqs.(4.1) reveal that such remote arrangements survive for non-infinitesimal stretch intensities in a range, and show that the numerical $x_k$ lie near the roots of Hermite polynomials even when their altitudes $|\Im(z_k)| \approx h$ are not large compared to $x_{max}$ any longer; see Table III.

| $\|k\|$ | Numerical $x_k$ s | Hermite $x_k$ s |
|---|---|---|
| 1 | 0.71748382 | 0.66649627 |
| 2 | 1.4684599 | 1.36644918 |
| 3 | 2.3196838 | 2.16531738 |

Table III: Comparison between the numerically determined abscissas $x_k > 0$ of horizontally aligned "remote" poles, $\pm x_k \pm ih$, with those estimated from the Hermite polynomial $\mathcal{H}_N(x_k(S/2\nu)^{1/2}) = 0$, for $N_1 = 7$, $\nu = 1/10$ and $S = 0.30$; the turning point is at $S^* \approx 0.69$, and numerics gives $|\Im(z_k)| \approx 2.53$ only.

Comparisons of the above near-horizontal remote pole arrangements with references [6] [30] suggest that a similar stretch-based mechanism underlies the 'interpolating solutions' of the MS equation evidenced there, where approximately horizontal arrangements of poles are found to lie at a distance above and between the vertically-aligned ones pertaining to the main crests; the stretch is then due to the main crest curvature $\Phi_{xx}$ and, *in fine*, results from the attraction by the array of vertically-aligned poles belonging to the space-periodic base front slope $\Phi_x$ (instead of the pole-at-infinity of $Sz$ in the present situation).

As if is not enough, for some given sets $\{S, \nu, N_{tot}\}$ either type of pole arrangement (all close to the real axis, *or* near the altitudes $\pm h \gg \nu N_k$) can be obtained, depending on the initial seed chosen in Newton iterations ; each type may have its own turning point, at least when $N$ is moderate .

The two types can even coexist in some instances, which contributes to a further proliferation of solutions. As shown in Fig.4, six different patterns are already allowed for the same stretch intensity (e.g. $S = 1$) when $N = 3$, not to mention those belonging to $N = 1, 2$ only. The situation does not simplify as $S\nu > 0$ gets smaller, since larger $N$s are allowed for. It is not excluded that the arrangements obtained so far are only the first members of an endless family, with a backbone combining nearly vertical and nearly horizontal pole arrangements in a hierarchical manner at larger and larger scales as $0 < S\nu \ll 1$ decreases. Two features complicate the matter (see Fig.9 for solutions with $N$ up to 8): (i) the couplings between the $x$- and the $B$- wise interactions (*e.g.*, see Fig.4 at $S \approx 0.4$); (ii) the existence of pole detachments from the main pile(s), which typically corresponds to the upper branches belonging to a given total pole number: *e.g.*, a near-real (1,1,1) Hermite type of solution becomes a (1,1$^+$,1) one as $0 < S\nu$ decreases along the upper branch with same turning point, or (1,1,1,1) becomes (1$^+$,1,1,1$^+$) ... .
To summarize Sec. VI: *weak positive stretch allows for very numerous novel steady solutions.*

## VII. SELF-SIMILAR EVOLUTIONS

### A. Nearly real poles

The steady patterns analysed in Sec.V are now shown to be special cases of analytically accessible evolutions. We firstly consider the situation of near-real pole arrangements analysed in the paragraphs below (5.2), yet with separated enough crest locations $X_k(t)$ that are now off the steady situations described therein, and hence obey unsteady 'attraction/expansion' balances

$$\frac{dX_k}{dt} = -\sum_{1=j \neq k}^{j=M} \frac{4N_j \nu}{X_k - X_j} + SX_k, \tag{6.1}$$

when $0 < S\nu N_k \ll 1$, $k=1,...,M$: the typical time-scale involved in (6.1) is $t = \mathcal{O}(1/S)$ and hence largely exceeds that, $\mathcal{O}(\nu N_k)$, needed for the $2N_k$ poles 'inside' the $k$-th crest to align nearly vertically at $\Re(z) \approx X_k$, $\Im(z) = \mathcal{O}(\nu N_k)$. We next invoke a polynomial $\Pi_M(\xi)$, the $M$ zeros of which $\xi = \xi_k$, $k=1,...,M$, are all assumed real and satisfy:

$$0 = -\sum_{1=j \neq k}^{j=M} \frac{N_j}{\xi_k - \xi_j} + N_1 \xi_k \quad . \tag{6.2}$$

For example, $\Pi_M(\xi)$ relates to Laguerre polynomials of $\xi^2$ in the most complicated example evoked in Sec.V; $\Pi_2(\xi)$ with $N_1 \neq N_2$ also is available there. A direct substitution shows that

$$X_k(t) = L(t)\xi_k + D(t), \quad k=1,...,M \quad , \tag{6.3}$$

are solutions of (6.1), provided $L(t)$ and $D(t)$ follow uncoupled differential equations:

$$dD/dt = SD, \tag{6.4}$$
$$dL/dt = -4\nu N_1/L + SL . \tag{6.5}$$

While $D(t)$ accounts for the already mentioned 'recession' of any pattern as a whole, the scale factor $L(t)$ controls all mutual x-wise distances between crests. For a constant $S$, (6.5) gives

$$L(t)^2 = (L^0_{eq})^2 + (L^2(0) - (L^0_{eq})^2)\exp(2St) ,\qquad(6.6)$$

with $L^0_{eq} \equiv (4\nu N_1 / S)^{1/2}$. By equation (6.6), *multi-crest nearly-real equilibrium configuration are unstable*, since $|L(0)| > L^0_{eq}$ ultimately leads to $|L| \sim \exp(St)$ and a uniform exponential stretching of all mutual distances between crests: 'expansion' ultimately wins over attraction. On the contrary, $|L(0)| < L^0_{eq}$ leads to the simultaneous collapse of *all* crests into a single one located at the abscissa $D(t_{merger}) = D(0)\{1-(L(0)/L^0_{eq})^2)\}^{1/2}$, $(1-(L(0)/L^0_{eq})^2)\exp(St_{merger}) \equiv 1$; actually, the 1-D approximate dynamics (6.1) ceases to be valid for $t \lesssim t_{merger}$ because the crests are not sufficiently separated any longer, and later. Numerical integrations of Eqs.(4.1) were performed with $M = 7$, $N_k = 1$ for all crests, $S = 1/100$, $\nu = 1/10$, and initial locations $z_k(0) = L(0)\xi_k \pm i\nu$, $L(0) \neq L^0_{eq}$, $D(0) = 0$. An agreement with (6.3)(6.6) was obtained, except for a slight discrepancy near the time $t_{merger}$ if $L(0) < L^0_{eq}$, when the dynamics of $L(t)$ ceases to be slow ($\max(|X_k|/\nu)$ gets too small) and the cumulated numerical errors are enough to break the self-similarity of the final collapse. Repeating the simulation with $z_k(0)$s that are randomly displaced by a few percents from $L(0)\xi_k \pm i\nu$ also led to fair agreement with (6.3)(6.6) when $L(0) > L^0_{eq}$ see Fig.5. Though with $L(0) < L^0_{eq}$ premature pair-wise coalescences occur instead of a single 7-pair crest being formed at once, see Fig.6 ; a single crest will form soon after, however, by the alignment mechanism recalled at the beginning of Sec. V.

### B. Remote poles

The above analysis can be adapted to the case of $N$ remote pole pairs that are initially nearly aligned horizontally at $z_k = x_k(0) \pm ih(0)$, $h(0) = \mathcal{O}(1/S) \gg \nu N$. To satisfy the imaginary part of (4.1), their current common altitude $h(t) \gg \mathcal{O}((N\nu/S)^{1/2})$ must obey

$$dh/dt = N\nu/h - 1 + Sh + ... \qquad(6.7)$$

instead of (5.4), whereas the unsteady version of (5.5) ($dx_k/dt$ is added to the right-hand side) has solutions $x_k(t) = L(t)\xi_k + D(t)$ still given by (6.2)(6.3)(6.6), up to a few differences.

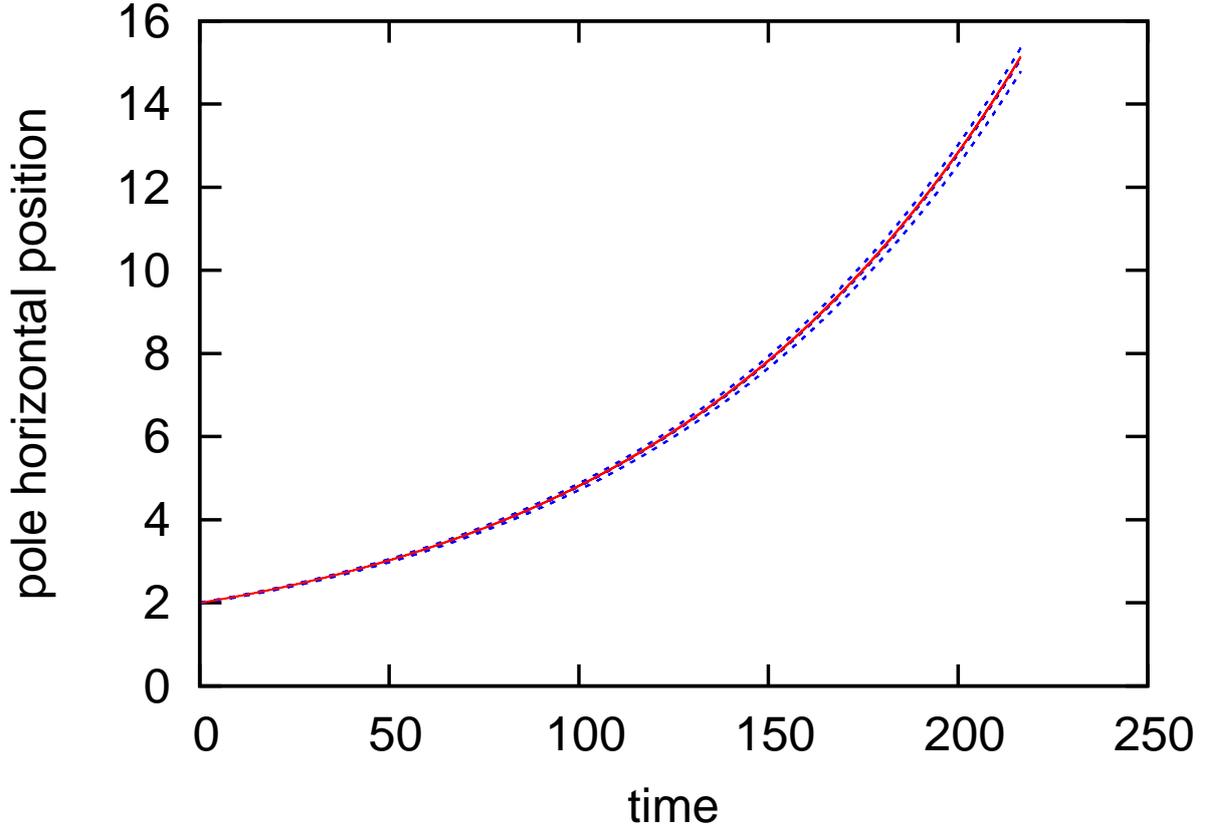

FIG. 5 (Color online) Dotted (blue) lines: Evolutions of the scaled abscissas $X_k(t)/L_{eq}^0 \xi_k > 0$ of 3 (out of $M = 7$) nearly real crests initially at about $X_k(0)/L_{eq}^0 \xi_k = 2$ (*twice* the equilibrium values), for $S = 1/100$, $\nu = 1/10$, $N_{k \geq 1} = N_1 = 1$, and purposely disturbed initial poles locations (see text). Solid (red) line: analytical prediction, Eq.(6.3).

(i) Firstly, one must set $N_k = N_1 = 1$ in (6.2)(6.5), whereby the equilibrium scale factor now is $L_{eq}^\infty \equiv (2\nu N/S)^{1/2}$. Yet like previously, *steady nearly-horizontal remote pole arrangements are unstable*: $|L| \sim \exp(St)$ for $St \gg 1$ if $|L(0)| > L_{eq}^\infty$, whereas $|L(0)| < L_{eq}^\infty$ leads to the collapse of all (simple-) poles into a single one (of order $N$) at the finite $t = t_{merger}$, $St_{merger} \equiv -\ln(1 - (L(0)/L_{eq}^\infty)^2)$. (ii) Another difference with near-real poles is that the dynamics (6.5) does not stop at $t = t_{merger}$. $L^2(t)$, as is now defined by (6.6), may get negative and $L(t)$ itself imaginary: the $N$ poles then become *vertically* aligned, and will remain nearly so for $t > t_{merger}$.

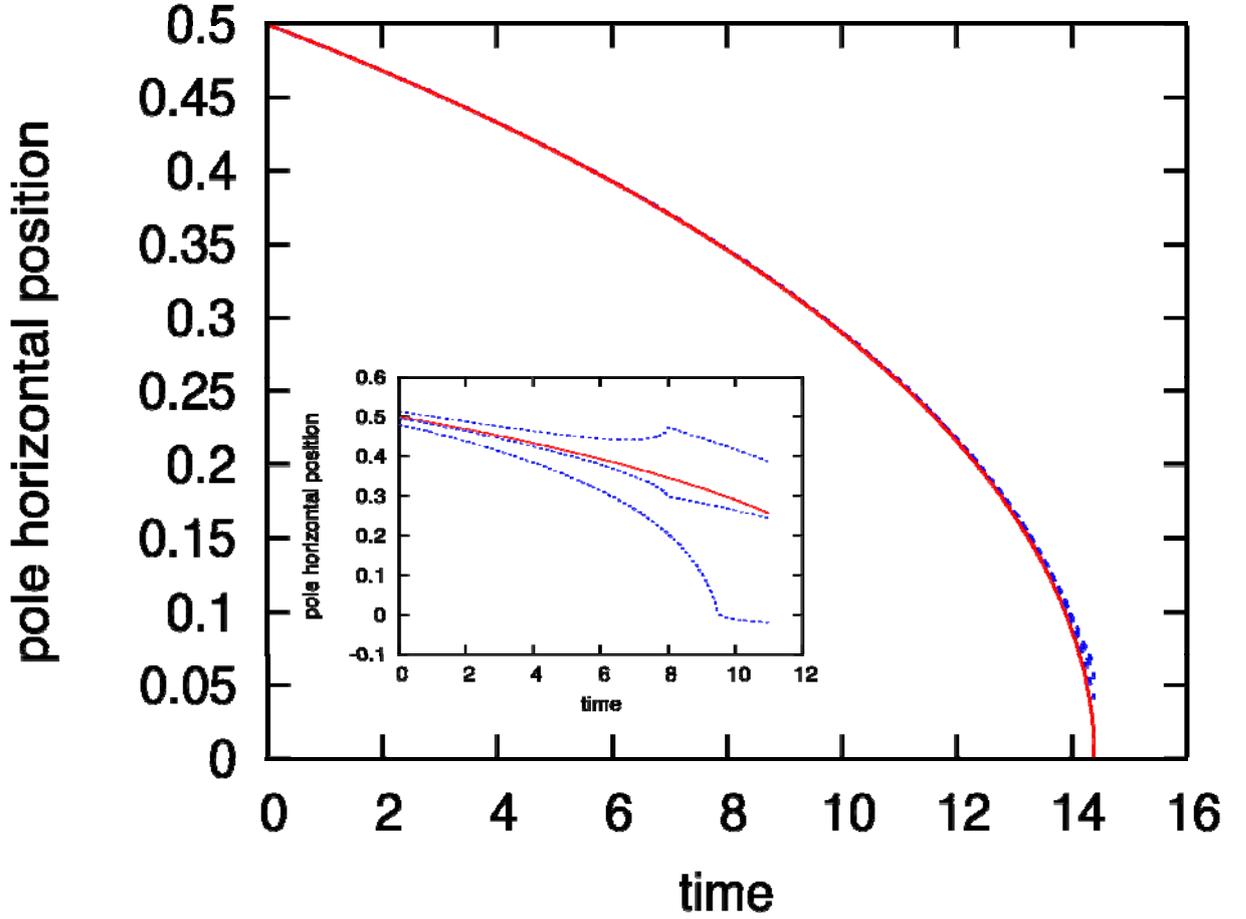

FIG.6 (Color online) Dotted (blue) lines: Evolution of the scaled abscissas $X_k(t)/L_{eq}^0 \xi_k > 0$ of 3 (out of $M = 7$) near-real crests initially at $X_k(0)/L_{eq}^0 \xi_k = 1/2$ (*half* the equilibrium values), for $S = 1/100$, $\nu = 1/10$, $N_{k \geq 1} = N_1 = 1$. Solid (red) curves: analytical prediction, Eq.(6.3). Inset: like previously, with the same intentional "noise" on initial conditions as in Fig.5.

The latter scenario of course assumes that the altitude $h(t)$ has not shrunk to zero in the interim: whereas $h(0) > h_{eq} = 1/S - \nu N + ...$ eventually leads to $h \sim \exp(St)$ and poles that leak to $\pm i\infty$, any $h(0) < h_{eq}$ will drive the solution of (6.7) to $h(t_0) = \mathcal{O}((N\nu/S)^{1/2}) \ll 1/S$ for $t \approx t_0$, $St_0 \equiv -\ln(1 - h(0)/h_{eq})$; after $t_0$ the poles in question will shortly become nearly real, $h(t > t_0) = \mathcal{O}(\nu) \ll (N\nu/S)^{1/2}$. The analysis just given then ceases to be valid, but the previous one pertaining to nearly real pole pairs becomes applicable: at $t \approx t_0$ the poles are indeed separated by $\mathcal{O}((N\nu/S)^{1/2})$ horizontal distances, and in first approximation for $0 < SN\nu \ll 1$ their abscissas are still proportional to the zeros $\xi_k$ of a Hermite polynomial, $\mathcal{H}_N(\xi)$. If $t_0 < t_{merger}$, *i.e.* $h(0)/h_{eq} < (L(0)/L_{eq}^\infty)^2$, and such that $L(t_0) < L_{eq}^0$ (or $L(t_0) > L_{eq}^0$) a $N$-crest pattern will be seen to crop on top of the base flame before collapsing into a single one (or a $N$-crest pattern expanding laterally); on the contrary, if $t_0 > t_{merger}$, only an isolated crest will be observed when acquiring a significant amplitude.

Thus, depending on the values of the initial altitude $h(0)$ and scale factor $L(0)$, a variety of behaviours caused by nonlinear interactions may take place among initially remote pole arrangements, even though this can hardly be noticed from the real axis because the poles involved are too far from it when these 'off-stage' processes occurs. Moreover, slight differences in the initial conditions may result in completely different patterns as time elapses: *the unstable equilibriums* (*e.g.*, remote poles at $\xi_k L_{eq}^\infty \pm i h_{eq}$, or near-real ones at $\xi_k L_{eq}^0 \pm i.\mathcal{O}(\nu N_k)$) *play the part of 'shunting-' or 'saddle-' points for the system trajectories*. Since the poles are indiscernible (identical residues) the existence of unstable equilibriums, whose number quickly increases as $1/S\nu \to +\infty$, almost precludes one from tracing back the origin of sub-wrinkles of a weakly curved flame front from the sole observation of their shape, location and amplitude when they get visible. This likely contributes to the nearly random manner sub-wrinkles crop up on top of weakly curved flame troughs [8], like in a Galton box.

## VIII. TIME-DEPENDENT STRETCH

A too intense constant stretch $S > 0$, for example induced by a too strong steady $u(x)$ in Eq.(2.1), can moderate or inhibit the phenomenon of trough-splitting (*i.e.*, crest formation) that the DL instability mechanism tends to induce. On the contrary, a constant compression, $S < 0$, tends to pull the poles of $\phi_x$ and to make them crowd near the origin $z = 0$. What happens when $S$ oscillates and possibly changes sign with time has so far not been investigated, even though this relates to the sub-wrinkles of flames subjected to a time-dependent, non-uniform $u(t,x)$ in Eq.(2.1). Oscillating stretch intensities $S(t) = \langle S \rangle + \sigma \sin(\omega t)$ of various mean values $\langle S \rangle$, amplitudes $\sigma$, and frequencies $\omega$ are considered below.

At frequencies $\omega \ll 1/\nu$ the quasi-steady approximation applies, whereby a slowly evolving 2-pole centred crest may not be durably viable whenever $\max(S) = \langle S \rangle + \sigma$ exceeds $S_c = 1/4\nu$, see Sec.III.A. Its quasi-steady structure will cease to exist when $S(t)$ crosses $S_c$ for the first time : even if $S(t)$ may later cross $S_c$ from above, $B_1(t)$ will have meanwhile moved high enough above the unstable $B^+(\nu, S(t))$ not to be again attracted near the lower branch $B^-(\nu, S(t))$ of quasi-steady solutions. Likewise, centred crests with $2N \gg 1$ vertically aligned poles are ruled out whenever $\langle S \rangle + \sigma$ exceeds $S^*$ for long enough, Eq.(3.8), but can nevertheless survive with a smaller number of pole pairs, $N \leq \lfloor 1/2\pi^2 \nu (\langle S \rangle + \sigma) \rfloor$ (see Sec. III.B). In other words, *at least if slow the $S(t)$ history selects $N$ in the long-time limit.*

To study the opposite, high-frequency, case $\omega \gg 1/\nu$ one first sets $B_n = \beta_n E(T)$, $E \equiv \exp(-(\sigma/\omega)\cos T)$, $T = \omega t$, and next invokes a two-time ($t$ and $T$) asymptotic method, formally replacing $d/dt$ by $\omega \partial_T + \partial_t$. Provided they remain $\mathcal{O}(N\nu)$ to leading-order for $\omega \gg 1/\nu$ the $\beta_n(t,T)$-functions may actually only depend on the 'slow' time $t$; taking an average (noted $\langle . \rangle$) of (4.1) over $T$ shows they obey the 'slow dynamics':

$$\frac{\partial \beta_n}{\partial t} = \sum_{-N=m \neq n}^{m=N} \frac{2\nu K}{\beta_n - \beta_m} - J.\text{sgn}(\beta_n) + \langle S \rangle \beta_n . \qquad (7.1)$$

Here $a \equiv \sigma/\omega$, $J \equiv \langle \exp(\pm a.\cos(\omega t)) \rangle = I_0(a)$ with $I_0(.)$ being the modified Bessel function of first kind and zeroth order, and $K \equiv I_0(2a)$. This averaged dynamics for the $\beta_n$s thus has the same structure as the restriction of (4.1) to aligned poles $z_n = iB_n$, up to coefficients that only depend on $(\sigma/\omega)$, *i.e.* on the power spectrum of the integrated stretch fluctuations.

In particular, the late time state of a two-pole ($N=1$) oscillating crest must satisfy $\nu K/\beta_1 - J + \langle S \rangle \beta_1 = 0$ instead of (3.3), and will be allowed only if $\langle S \rangle$ is less than $\langle S \rangle_c$, with

$$\langle S \rangle_c = S_c J^2 / K < S_c, \qquad (7.2)$$

where $S_c = 1/4\nu$ is the same as in the non-oscillating, two-pole case, see (3.3); the corresponding value of $\beta_1$ is $2\nu K/J > B_c = 2\nu$; that of $\langle B_1 \rangle$ is even larger, $\langle B_1 \rangle = 2\nu K$, not to mention $\max(B_1(t)) = \beta_1^\pm \exp(a) > \nu K \exp(a)$. In other words, an intense enough high-frequency oscillating component of the stretch intensity tends to flatten a two-pole crest ($\langle \beta_1 \rangle > B^-$ along the lower branch $\beta_1^-$ of steady solutions to (7.1)): more importantly, fast enough *oscillations in* *stretch intensity make sub-wrinkle suppression easier than by the mean stretch alone*, $S_c \geq \langle S \rangle_c$, yet moderately so since $J^2/K$ only slowly decays from $1 - a^2/2 + ...$ at $|a| \ll 1$ to $\approx 1/(\pi |a|)^{1/2}$ at $|a| \gg 1$. One may also note that $2\langle S \rangle_c \langle B \rangle_c = J > 1$, indicating a weaker and wider crest profile at turning point than without fluctuations of stretch intensity.

The numerical integration of (4.1) with $N = 2$, $\nu = 1/10$, $\omega = 20$, $a = 3/2$ ($J^2/K = 0.555...$) shows that a successful two-time analysis does not require $\omega$ to largely exceed the reciprocal relaxation time $t_1$ of $\beta_1(t)$ to the steady root $\beta_1$: in Fig.7, $\omega t_1 \approx 3.5$. The large-$\omega$ predictions tend to be more accurate for the highest poles.

Likewise, crests with $N \gg 1$ aligned pole pairs subjected to high-frequency stretch are ultimately described by an equation similar to (3.4), yet with $\{\nu, S\}$ replaced by $\{\nu K/J, \langle S \rangle/J\}$. They may therefore exist only if $2\pi^2 N \nu \langle S \rangle < J^2/K < 1$: again, the condition for stretch-induced crest suppression is more restrictive than (3.8). In the large-$\omega$ limit the instantaneous crest slope has $E(T)\phi_x = -\int_0^{+\beta_{max}} 4\nu X p(\beta) d\beta/(X^2 + \beta^2)$, with $X \equiv x/E(t)$, and the equations that determine the $p(\beta)$ have the same structure as (3.4) provided $\{\nu, S, B, B_{max}\}$ are replaced therein by $\{\nu K/J, \langle S \rangle/J, \beta, \beta_{max} = \langle B_{max} \rangle/J\}$; the corresponding instantaneous crest shape is ultimately self-similar, $\phi(t, x) = F(X)$.

Before closing this section, a few remarks may be put forward.

The large-$\omega$ analysis can be easily adapted when the fluctuation $S'(t)$ of stretch intensity contains $R$ widely separated frequencies $1/\nu \ll \omega_1 \ll \omega_2 \ll ... \ll \omega_R$, with partial amplitudes $\sigma_r$: the influence of each frequency can be accounted for in its turn, starting from $\omega_R$. As a result, the $J$ coefficient involved in the slowest dynamics (7.1) becomes $J = I_0(a_1) I_0(a_2) ... I_0(a_R)$, $a_r \equiv \sigma_r/\omega_r$, and $K$ acquires a similar expression. Because $\ln(I_0(|a| \ll 1)) = a^2/4 + ...$, such products converge for $R = \infty$ whenever the power spectrum of the integrated stretch fluctuation satisfies the comparatively mild condition $\sigma_r/\omega_r = o(1/r^{1/2})$. Whereas also allowing $\langle S \rangle$ to depend on slow time $t$ is a harmless further generalization, it is not known whether the above elementary 'cascade renormalization' [31] can be extended to a continuous spectrum $\omega \geq \omega_1 \gg 1$ (replacing the series for $\ln J$ and $\ln K$ by integrals over $\omega \geq \omega_1$), and to turbulent-like fluctuations of $S$; the analogues of $J$ and $K$ could even be tabulated numerically, if $S'(t) = S(t) - \langle S \rangle$ is available and contains no beat with $\omega \nu = \mathcal{O}(1)$.

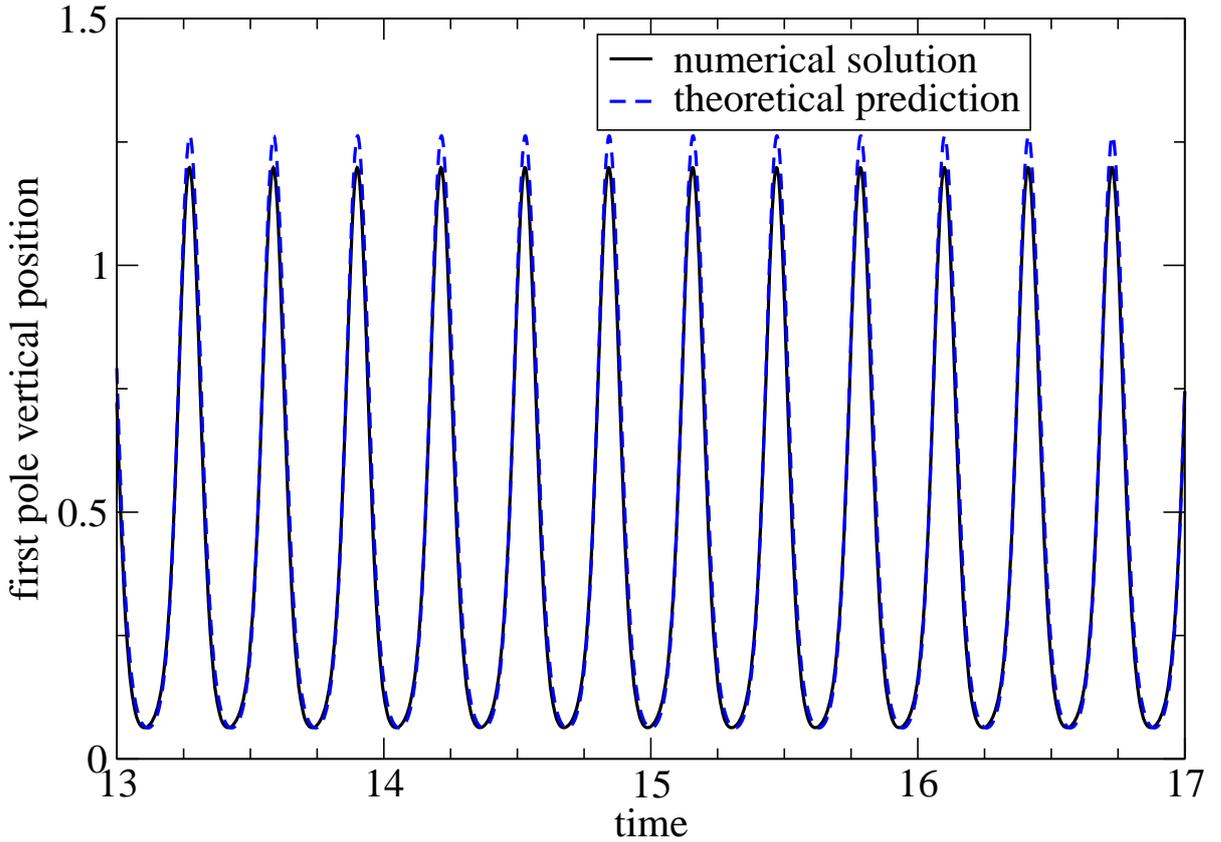

FIG.7 (Color online) Black solid line: oscillating solution $B_1^-(t)$ obtained from numerical integration of equation (4.1) for $N = 2$ ($z_k = iB_k$, $|k| = 1, 2$) for $\nu = 1/10$, $S(t) = 0.2 + 30\sin(20t)$. Dashed (blue) line: analytical curve $\beta_1 \exp(-(\sigma/\omega)\cos T)$, $\beta_{1,2}$ being steady solutions of (7.1).

At any rate, the convexity of exponential functions guarantees that $\langle E \rangle \geq 1$, $\langle E^2 \rangle \geq 1$ for any $d(\ln E)/dt = S'(t)$ such that the latter time-averages exist. The Cauchy-Schwartz inequality implies $J^2/K = \langle E \rangle^2 / \langle E^2 \rangle \leq 1$, whereby *the trends revealed with harmonic variations of stretch intensity are fairly general*. They actually extend to fronts that are $2\pi E(t)/\kappa$-periodic ($\kappa > 0$) in $x$, in which case $S'(t) = d(\ln E)/dt$. This pertains to $x \to -x$ invariant flame fronts propagating along the centreline ($y$-axis) of left/right symmetric two-dimensional channels with a variable width $\Lambda(y) > 0$ [32]. The scale factor is then $E(t) = \Lambda(\hat{t}.u_L)/\langle\Lambda\rangle$, $\hat{t} \sim t\ell/u_L\mathcal{A}^2$ being a dimensioned time, whereby $\langle S \rangle = 0$ and $\langle E \rangle = 1$; the analogue of $J/K$ again is less than 1. One can show that the number $N$ of pole pairs present 'in' the channel must be less than $\lfloor (1 + J/\nu\kappa K)/2 \rfloor$ if a cellular pattern is to survive; this is more stringent than when $\Lambda(y) \equiv \langle\Lambda\rangle$ (i.e., $J = K$), and no wrinkle is allowed durably in the wavy channel if $J/K \leq \kappa\nu$.

## VIII. STRETCH *vs.* SPECTRAL CUTOFF

To put the findings obtained so far in a more physical perspective, and relate them to the second problem evoked in Sec. **I**, it is useful to restore dimensions in the 'stretched' MS equation (2.3). This gives:

$$\hat{\phi}_{\hat{t}} + a(\mathcal{A})u_L[\Sigma.\hat{x}\hat{\phi}_{\hat{x}} + \tfrac{1}{2}\hat{\phi}_{\hat{x}}{}^2] = u_L\Omega(\mathcal{A})[\hat{\phi}_{\hat{x}\hat{x}}/k_n + I(\hat{\phi},\hat{x})], \qquad (8.1)$$

where the over-hats of $\hat{x}$, $\hat{t}$, ...denote dimensioned variables, and the uniform drift velocity $\hat{U}(\hat{t}) \sim U(t)$ is again omitted: $\hat{\Phi} \equiv \Sigma.\hat{x}^2/2$ represents the parabolic background front, on top of which sub-wrinkles of local amplitude $\hat{\phi}(\hat{t},\hat{x})$ grow. The dimensionless grouping $a(\mathcal{A}) \approx (1+\mathcal{A})$ and the DL coefficient $0 \leq \Omega(\mathcal{A}) \approx ((1+\mathcal{A})/(1-\mathcal{A}))^{1/2} - 1$ only depend on the Attwood number $0 < \mathcal{A} = (\rho_u - \rho_b)/(\rho_u + \rho_b) < 1$, and are known from separate analyses [1]-[4].

The 'ultimate' steady sub-wrinkle (in the proper frame) that can survive as the stretch intensity $u_L\Sigma$ (now a reciprocal time) increases is the *two*-pole solution at its turning point (the same as at $S = S_c = 1/4\nu$ in (3.3)), see Figs 4 and 9. It corresponds to a total (*i.e.*, base-flame + sub-wrinkle) dimensioned flame profile $\hat{\varphi} \equiv \hat{\Phi} + \hat{\phi}$ of the form:

$$\hat{\varphi}_c = \frac{\Omega(\mathcal{A})}{a(\mathcal{A})k_n}\psi(\hat{x}k_n), \quad \psi(\xi) \equiv \frac{\xi^2}{8} - 2\ln(1 + \frac{\xi^2}{4}). \qquad (8.2)$$

The function $(3/4 - \ln 4)(1 - \cos(\pi\xi/\sqrt{12}))$ accurately osculates $\psi(\xi)$ at and between its min or max, located at $\xi = 0, \pm\sqrt{12}$, see Fig.8. This suggests that, as a result of the geometrical stretch caused by distortions at larger scale, *visible wrinkles with wave-numbers* $|k| \geq k_{cutoff} = (\pi/2\sqrt{3})k_n \approx 0.9k_n$ *unlikely exist*, provided the local wrinkle structure be considered quasi-steady. Since the crest radii of curvature essentially scales like $1/k_n$ regardless of their amplitude [7], this result is compatible with the experimentally known near-equality between neutral and cut-off wave-numbers for the wrinkle spectrum of flames propagating in turbulent flows.

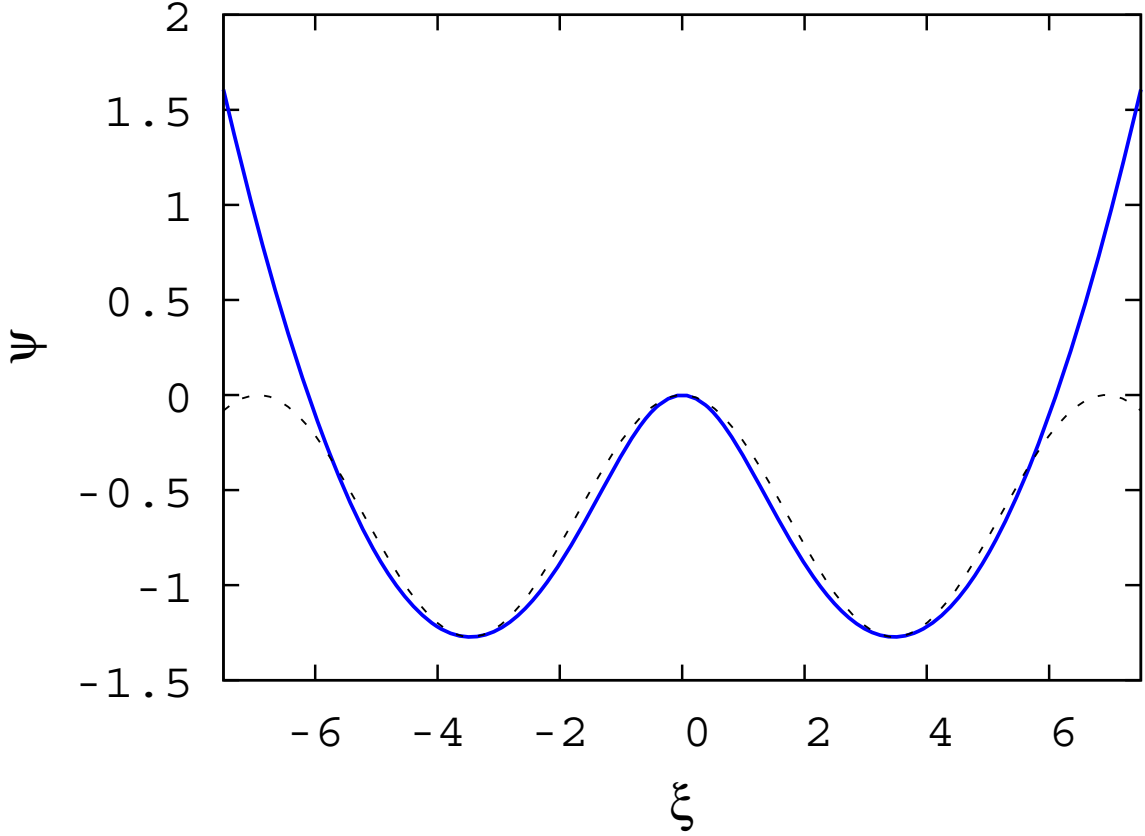

FIG.8 (Color online) Solid (blue) curve: total flame front shape at turning point $\Sigma = \Sigma_{cutoff}$ of the two-pole crest, $\psi(\xi) \equiv \xi^2/8 - 2\ln(1+\xi^2/4)$, vs. $\xi = \hat{x}\hat{k}_n$. Dashed (black) line: osculating cosine (see below Eq.(8.2) ).

This result $k_{cutoff} \approx k_n$ is not a mere dimensional consequence of having the Markstein length $\mathcal{L}$ as reference in problems of flame dynamics, since $k_n\mathcal{L}$ depends on the Attwood number: to wit, $k_n\mathcal{L} \approx \Omega(\mathcal{A}) \sim \mathcal{A}$ for $\mathcal{A} \ll 1$. Next, the condition for suppression of all steady sub-wrinkle by geometrical stretch ($S > S_c$ in the notation of Sec.III) can be re-written as $\mathcal{D} > \mathcal{D}_c \equiv 1/a$, where $\mathcal{D}$ is a Damköhler number [33]. It is defined as the ratio of the rate-of-strain $u_L\Sigma$ based on flat-flame speed $u_L$ and the background flame curvature $\Sigma$ (tributary of the forcing function in (2.1) ), to the maximum DL growth rate $\Omega(\mathcal{A})u_Lk_n$; this suggests a dynamical origin. That *a balance of nonlinearity, DL instability and stretch is a sine qua non of wrinkle suppression* (see (3.3)) confirms the genuine *flame-dynamical origin* of the $\mathcal{D} > \mathcal{D}_c$ criterion, and the presence of $a(\mathcal{A})$ (weight of nonlinearity in (8.1)) in $\mathcal{D}_c$ further substantiates this.

The unsteadiness caused by a time-dependent $\Sigma$ will admittedly bring about numerical factors (*e.g.*, the grouping $J^2/K$ in (7.2)) in the above criterion; yet those will stay $\mathcal{O}(1)$ unless $u_L \Sigma(\hat{t})$ *and* the frequencies $\omega$ it involves are well higher than $\mathcal{O}(u_L \Omega(\mathcal{A}) k_n)$. That, however, is unlikely for turbulent flames propagating through actual reactive gaseous pre-mixtures in conditions when a thin front of $\mathcal{O}(\ell)$ thickness can be identified as such, for this requires $u_L \Sigma(\hat{t})$ and $\omega$ to be well smaller than $u_L/\ell$ [33], and $\Omega(\mathcal{A}) = \mathcal{O}(1)$ in practice: the estimate $k_{cutoff} = \mathcal{O}(k_n)$ is then expected to still hold true in such regimes, which it does [10][34].

As shown in Section III, the approximate result $2S \max(B) \approx 1$ was obtained at turning points regardless of the number $N$ of pole pairs involved, thereby suggesting that $\mathcal{O}(1/\Sigma)$ typical wrinkle size (amplitude and width) could be selected by the fluctuating stretch intensity $u_L \Sigma(\hat{t})$, when $\Sigma \ll \Sigma_{cutoff} \equiv \Omega k_n / 4a$. This militates in favour of a (mean-) power spectrum of wrinkling tied to that of stretch intensity at moderately small wave-number ratios $k/k_n$: the estimate $\hat{B} \approx 1/2\Sigma$ for the typical sub-wrinkle amplitude $\hat{B}$ prevailing at larger scales than $1/k_{cutoff}$ might then form the basis of a scaling-law of the form $\Sigma^{-1}(k) f(k/k_{cutoff})$ for the (mean-) power spectrum of wrinkling, with a nearly constant (or very small) $f(.)$ at $k/k_n \ll 1$ (or $k/k_n \gg 1$).

## IX. FINAL REMARKS, OPEN PROBLEMS

Combining analytical and numerical approaches based on the pole decomposition, this work revealed that inclusion of geometrical stretch markedly modifies the otherwise simple [7, 19] isolated solutions of the classical (Michelson-) Sivashinsky PDE for (weakly-) unstable flames.

Firstly, it was demonstrated that accounting for a uniform stretch intensity $S$ is enough to generate novel types of isolated solutions. New steady (unstable-) solutions with arrays of horizontally aligned near-real or remote poles (or both) were evidenced; when $S = 0$ only *one* centred isolated solution, involving a single vertical pole alignment, existed whatever $N \geq 1$ is [21]. The net result is a proliferation of equilibrium front shapes, especially at small $S$s, Fig. 9.

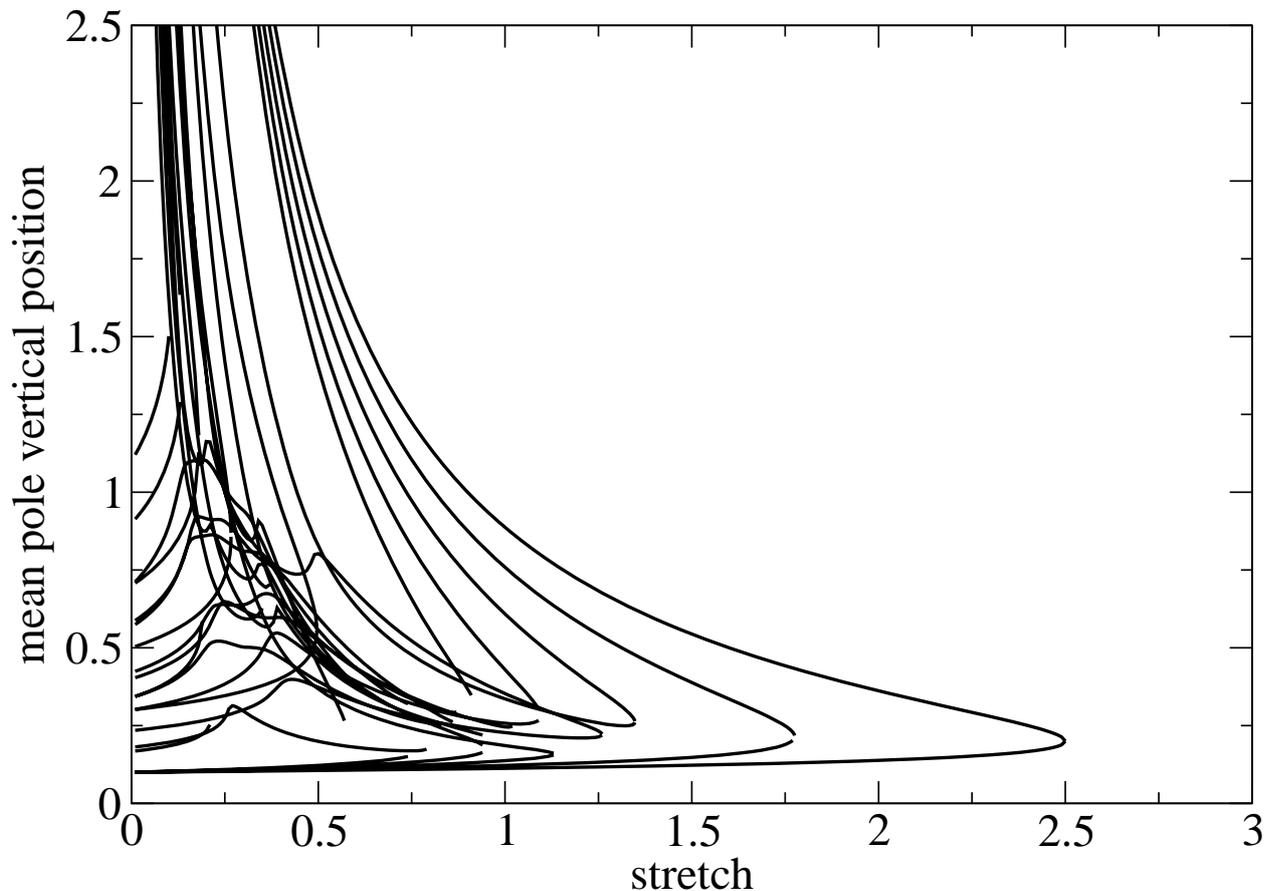

FIG. 9: Compiled $B_{bar}$ (upper-pole barycentre) *vs.* $S \geq 0$ (dimensionless stretch intensity) curves obtained in the present work for $v = 1/10$, up to $N = 8$, for some of them. Curves with $N \leq 3$ already appeared in Fig.4, e.g., the rightmost solution (3.3) ($N = 1$).

Part of the complexity of MS-type equations (even if $S = 0$) boils down to the fact, caused by a nonlinearity and a curvature term already present in the Burgers equation, that the spatial poles of $\phi_x$ are indiscernible and interact pair-wise *via* a $1/z$-law: midway between to two such singularities, another one can stay in equilibrium. A specificity of this nonlinearity-induced phenomenon is the anisotropy (horizontal attraction *vs.* vertical repulsion) of the $1/z = \overline{z}/|z|^2$ law. By contrast, stretch effects acts on front singularities (hence on its shape) in an isotropic way. Its presence brings about the possibility of new " stretch *vs.* horizontal attraction" and/or " stretch *vs.* Darrieus-Landau instability" partial equilibriums, thereby noticeably contributing to the quick proliferation of (unstable-) steady solutions when $S > 0$ decreases. A similar mechanism operates in periodic solutions of the stretch-free MS equation: $S$ is then effectively provided by the curvature of the main cell trough, and the many poles associated with its crests provide the means to pull extra wrinkles/poles away from the trough, in a nearly isotropic way.

The analogy is further substantiated by the fact that the new stretch-induced 'steady' states found above are unstable, like those evidenced in [6][30]; their self-similar evolution, described analytically, will likely help handle those bursts travelling along wide front troughs.

It has also been shown that too intense stretch suppresses all isolated 'steady' wrinkles (Fig. 9), the larger/wider the easier: this is not caused by a local quenching of combustion processes inside the front structure [33], but results from an untenable balance between stretch, geometrical nonlinearity and nonlocal hydrodynamics (and curvature) . As the above analyses showed it, this effect is only quantitatively modified when the stretch intensity oscillates, crest suppression being just made somewhat easier. These findings have been used to suggest this mechanism as the reason why the wrinkle upper cut-off wave-number in turbulent flames experimentally coincides [10] with the neutral wave-number identified by linear stability analyses.

Classical (orthogonal) polynomials were encountered when studying stretch-induced horizontal pole/crest equilibriums. Not unduly surprising, for these are not directly affected by the Darrieus-Landau mechanism, and hence are electrostatic-like. The needed polynomials obey differential *local* equations, yet the classical ones do not cover all 'horizontal' equilibriums at weak stretch; some available generalisations (*e.g.* Heine-Stieltjes polynomials [35]) will hopefully do.
But concerning the 'vertical' equilibriums the situation is much less clear. The *nonlocal* DL instability mechanism indeed acts on the pole population in a way that is explicitly 'vertical', encoded as it is in an irremovable Hilbert transform. Further studies of the "stretched Sivashinsky polynomials", whose roots obey the steady pole equations, seem warranted: *e.g.*, to theoretically elucidate the very *nature* of the miracle that allows the MS equation to have pole-decomposed solutions (hidden symmetries or mere good fortune?), and because the Darrieus-Landau mechanism is an indispensable ingredient of stretch-induced wrinkle suppression.

Yet another important theoretical point is dangling. Implicit in the discussion about the influence of stretch on the wrinkle wave-number (Sec.VIII) was the assumption that the maximum admissible number of front-slope poles, corresponding to turning-point conditions, is the relevant one: since the pole dynamics conserves their number (when finite), how can noise supply 'enough of' them as the stretch intensity varies? This brings one back to a *nonlinear* problem already evoked in Sec.I: how does noise, even if weak when seen on the real axis, implant complex poles (incipient wrinkles)? Answering this question is one of the most challenging open theoretical issues about flame dynamics, not to mention its statistical aspects.

## APPENDIX A: THEORETICAL POLE DENSITY

Setting $B = B_{\max} \sin\theta$, $\text{sgn}(B) = \text{sgn}(\theta)$ and $P(B) = P(-B)$ are expanded as Fourier series:

$$\pi \,\text{sgn}(B) = \sum_{j=0}^{\infty} 4\sin((2j+1)\theta)/(2j+1), \tag{A.1}$$

$$P(B) = \sum_{j=0}^{\infty} P_j \cos((2j+1)\theta). \tag{A.2}$$

Employing them in the identity [21]

$$\unicode{x2A0F}_{-\pi/2}^{+\pi/2} \frac{\cos((2j+1)\theta')\cos\theta'}{\sin\theta - \sin\theta'} d\theta' = \pi \sin((2j+1)\theta), \quad j = 0,1,..., \tag{A.3}$$

then in (3.4) yields $2\pi^2 \nu P_j = 4/(2j+1) - \pi S B_{\max} \delta_{0j}$ ($\delta_{0j}$ = Kronecker delta); summing the series for $P(B)$ gives equation (3.5) of the main text.

## APPENDIX B : STIELTJES TRICK FOR CREST LOCATIONS

Multiplying each (5.3) by $X_k = -X_{-k}$, and setting $X_k = \text{sgn}(k)(4\nu N_1 \eta_{|k|}/S)^{1/2}$ leads to

$$2\eta_k \sum_{1=j \ne k}^{N} \frac{1}{\eta_k - \eta_j} + \alpha + 1 - \eta_k = 0, \tag{B.1}$$

with $\alpha = N_0/N_1 - 1/2$. As first realized by Stieltjes [29,35], the above sum over $j \ne k$ can also be written as $\lim_{\eta \to \eta_k} \{p'(\eta)/p(\eta) - 1/(\eta - \eta_k)\} = p''(\eta_k)/2p'(\eta_k)$, where $p(\eta) \equiv \Pi_{k=1}^{N}(\eta - \eta_k)$ and $(.)' = d(.)/d\eta$. Next, the polynomial $\eta p''(\eta) + (\alpha + 1 - \eta)p'(\eta)$ has $-N\eta^N$ as term of highest degree; according to equations (B.1) it vanishes when $\eta = \eta_k$, $k = 1,...,N$, and thus is $-Np(\eta)$. As the only polynomial solution of $\eta p'' + (\alpha + 1 - \eta)p' + Np = 0$ is an associated Laguerre polynomial, one has $p(\eta) \sim \mathcal{L}_N^{(\alpha)}(\eta)$ [24] and its accessible roots $\eta_k > 0$ give the crest locations, see Sec.V. The sum $1/\eta_1 + ... + 1/\eta_N$ of the roots $1/\eta_k$ of $\mathcal{L}_N^{(\alpha)}(\eta)/\eta^N$ is deducible from the known coefficients [24] of $\mathcal{L}_N^{(\alpha)}(.)$, and reads $2N/(\alpha+1) = 2N/(N_0/N_1 + 1/2)$.